\documentclass[
 aip,
 pop,
 amsmath,amssymb,
 reprint
]{revtex4-1}

\usepackage[utf8]{inputenc}
\usepackage[T1]{fontenc}
\usepackage{graphicx}
\usepackage{bm}
\usepackage{xcolor}
\usepackage{appendix}
\usepackage[final]{changes}

\providecommand{\az}{a(z)}  
\providecommand{\as}{a}  
\providecommand{\Bb}{B_b}
\providecommand{\nb}{n_b}
\providecommand{\Tb}{T_b}
\providecommand{\Es}{\textbf{E}_s}
\providecommand{\phis}{\phi_s}
\providecommand{\Hg}{H}
\providecommand{\HT}{H_T}
\providecommand{\HTo}{H_{T_0}}
\providecommand{\aT}{a_T}   
\providecommand{\aTo}{a_{T_0}}   
\providecommand{\hot}{h} 
\providecommand{\Mhot}{{{\rm M},h}} 

\usepackage{xcolor}

\begin{document}

\title{Collisionless stationary states of a stratified plasma in an expanding magnetic tube with stochastic heating}

\author{Luca Barbieri}
\email{luca.barbieri@obspm.fr}
\affiliation{LIRA, Observatoire de Paris, Universit\'e PSL, Sorbonne Universit\'e, Universit\'e Paris Cit\'e, CY Cergy Paris Universit\'e, CNRS, 92190 Meudon, France}

\author{Pascal D\'emoulin}
\affiliation{LIRA, Observatoire de Paris, Universit\'e PSL, Sorbonne Universit\'e, Universit\'e Paris Cit\'e, CY Cergy Paris Universit\'e, CNRS, 92190 Meudon, France}

\author{Daniel Verscharen}
\affiliation{Mullard Space Science Laboratory, University College London, Dorking RH5 6NT, United Kingdom}
\affiliation{Ruhr University Bochum, Faculty of Physics and Astronomy, Theoretical Physics I, Theoretical Plasma Physics, Bochum, Germany}

\date{\today}

\begin{abstract}
We investigate the collisionless kinetic structure of the upper solar atmosphere in the presence of an expanding magnetic field. Building on established velocity-filtration models, we extend the stochastic multi-temperature framework developed in previous works by including magnetic-field expansion and magnetic-moment conservation. In this framework, stochastic heating generates a non-Maxwellian boundary
distribution through the superposition of particle populations associated
with heating events characterized by different temperatures.

We consider a stationary two-component plasma confined within an expanding magnetic flux tube and subject to gravity, self-consistent electrostatic interactions, the Pannekoek--Rosseland electric field, and magnetic-moment conservation.
Starting from the Vlasov equation, we derive fully analytical expressions for the particle distribution functions, density profiles, and the parallel, perpendicular, and total temperature profiles.

We show that the combined conservation of energy and magnetic moment generates a loss-cone distribution, reducing the density with respect to the corresponding unmagnetized atmosphere and producing a pronounced temperature anisotropy. For a single-temperature boundary condition, the competition between magnetic-moment conservation and gravity causes the parallel temperature to develop a maximum. We derive analytical scaling laws for its location and amplitude and validate them against numerical calculations. In the stochastic multi-temperature case, gravitational filtering enhances the contribution of the hottest particle populations at coronal heights, while magnetic-moment conservation further amplifies the resulting velocity-space
anisotropy.

Our results provide a fully analytical kinetic description of the
combined effects of stochastic heating, gravitational filtering, and
magnetic-moment conservation in an expanding coronal magnetic flux tube,
and establish a collisionless benchmark for future kinetic models
incorporating more realistic magnetic-field geometries, such as dipolar
configurations, Coulomb collisions, and turbulent particle scattering.

\end{abstract}

\maketitle

\section{Introduction}
The atmospheres of the Sun and other low-mass main-sequence stars, together with stellar winds, constitute one of the most important examples of weakly collisional magnetized plasmas far from thermodynamic equilibrium. Although the particle mean free path in the
solar atmosphere is generally smaller than the characteristic density and temperature scale heights, it is not sufficiently short to ensure the establishment of local thermodynamic equilibrium. As a consequence, significant departures from local Maxwellian velocity distributions are observed both in the solar atmosphere \citep{Dudik_2017} and in the solar wind \citep{Maksimovic_al_2020}. Kinetic effects therefore play a fundamental role in determining the particle velocity distributions, density profiles, and thermodynamic structure of the corona and the solar wind
\citep{Chamberlain1960,Brandt1966,Jockers1970,Lemaire1971,Scudder1992a,Scudder1992b,Pierrard1996,Maksimovic1997,Landi-Pantellini2001,Vocks2001,Vocks_2002a,Vocks_2002b,Marsch2003,Vocks2003,Landi2003,Zouganelis2004,Vocks_2005,Marsch2006,Pierrard2011,Verscharen2019,Vocks2021,Vocks2021b,Jeong_2022,Barbieri2023temperature,barbieri2024temperaturedensityprofilescorona,Hau_2025,PtersdeBonhome2025,Barbieri2025c,Banik2026,Vinogradov2026}.

In  this paper, we illustrate the analytical results with an application to the solar atmosphere since it is the best observed among stars. Within the narrow transition region, only a few hundred kilometers thick, the temperature increases from approximately \(10^4\,\mathrm{K}\) in the upper chromosphere to about \(10^6\,\mathrm{K}\) in the low corona \citep{GolubPasachoff:book,observedtemperature}. Explaining the origin of this rapid temperature increase with height remains one of the fundamental unsolved problems in solar physics and is commonly referred to as the coronal heating problem
\citep{Klimchuk_2006,Reale2010,Klimchuk2015,VanDoorsselaere2020}.

The present work focuses on kinetic models of stellar atmospheres based on gravitational filtering
\citep{Scudder1992a,Scudder1992b,Meyer-Vernet_2007,
Barbieri2023temperature,Barbieri2024b,
barbieri2024temperaturedensityprofilescorona,Hau_2025,
Mullan_2025,Barbieri2025c,Banik2026,barbieri2026c}. This mechanism, also known as velocity filtration, arises from the preferential selection of high-energy particles as they move upward against gravity. Faster particles can reach greater heights in the gravitational potential than slower ones. Consequently, if the velocity distribution at the lower boundary contains a suprathermal tail, the relative contribution of high-energy particles progressively increases with height, leading to an increase in temperature with altitude. Motivated by observational evidence of spatially localized and temporally intermittent heating events in the upper chromosphere and transition region \citep{Young2018-sp,Berghmans:2021wl,narang2025,Harra2025,parenti2025}, recent studies have introduced a sequence of heating events in which the temperature increase is randomly sampled from a prescribed statistical distribution \citep{Barbieri2023temperature,Barbieri2024b}.  This is called stochastic heating. The term ``stochastic'' refers to the probabilistic sampling, which is used to mimic the variability of the heating events. This mechanism generates multi-temperature particle distributions near the chromosphere top.  At larger heights, these distributions are subject to gravitational filtering.  Recent studies show that this approach permit to reproduce the observed density and temperature structure of the upper solar atmosphere, so the transition region and the overlying million-degree corona \citep{Barbieri2023temperature,Barbieri2025c}.

However, these models \citep{Barbieri2023temperature,Barbieri2024b,Barbieri2025c} neglect the expansion of the coronal magnetic field. Indeed, in the photosphere the magnetic field is concentrated into mostly isolated magnetic flux tubes. As the plasma \(\beta\) decreases rapidly with height \citep{Gary2001}, the cross-sectional area of these flux tubes expands in order to maintain transverse force balance \citep{Gabriel1976,Dowdy1986,Solanki_1991,Judge_2021}. This expansion continues throughout the corona, although it progressively slows as
neighbouring flux tubes interact and fill the available coronal volume. This flux tube expansion leads to a gradual decrease of the magnetic-field strength with height \citep[e.g.][]{Mandrini_2000}. The resulting conservation of the first adiabatic invariant restricts the particle trajectories that remain magnetically connected to the lower boundary and therefore modifies the velocity-space distribution generated by gravitational filtering.

In the present work, we extend the stochastic-heating model by incorporating the expansion of the coronal magnetic field through a magnetic-flux-tube geometry used previously to model the magnetic field topology of the coronal field \citep[e.g.][]{Demoulin_1994}. Our goal is to investigate how the combined action of gravitational filtering and magnetic moment conservation modifies the density and temperature structure of the upper atmosphere.

As a first step, in this study we consider the collisionless limit and derive fully analytical expressions for the density, the parallel and perpendicular temperature profiles, and the total temperature. This allows us to identify the respective roles of gravitational filtering and magnetic field in determining the thermodynamic structure of the plasma and to isolate the purely kinetic effects associated with magnetic-field expansion.

The paper is organized as follows. In Sect.~\ref{sec:model_description}, we introduce the kinetic model of a collisionless plasma atmosphere embedded in an expanding magnetic flux tube. In Sect.~\ref{sec:single-temperature}, we derive analytical expressions for the density and temperature profiles corresponding to a single Maxwellian velocity distribution set at the lower boundary.  We investigate their general properties, and discuss the role of magnetic moment conservation in producing temperature anisotropies. In Sect.~\ref{sec:multi-temperature}, we move to solar application and therefore we extend the analysis to multi-temperature boundary distributions generated by stochastic heating, derive the corresponding analytical solutions, and investigate how magnetic moment conservation modifies the resulting density and temperature profiles. Finally, in Sect.~\ref{sec:conclusions}, we summarize the main results, discuss their implications for kinetic models of stellar atmospheres, compare with established fluid models \cite{Klimchuk_2006,Klimchuk2015} and original gravitational filtering approaches \cite{Scudder1992a,Scudder1992b,Meyer-Vernet_2007}, outline the limitations of the present approach, and suggest possible directions for future work.

\section{The model description}
\label{sec:model_description}
\subsection{Basic equations} \label{sec:Basic_equations}

We consider a plasma atmosphere composed of electrons and protons that is globally at rest. The atmosphere is embedded in a closed, loop-like magnetic field that prevents the formation of a stellar wind. We further assume a symmetric magnetic loop, with equal plasma pressure at both footpoints, so that no mass flow develops along the loop.

To simplify the analysis, we consider only a vertical magnetic field. The model therefore describes the nearly vertical legs of sufficiently large magnetic loops, whose height exceeds the gravitational
scale height of the atmosphere. This approximation avoids introducing a curvilinear coordinate along the loop, thereby reducing the complexity of the equations while retaining the essential physical ingredients.

Each particle is subjected to gravity,
\begin{equation}
\mathbf{F}_{g,\alpha}
=
m_{\alpha}\mathbf{g},
\qquad
\alpha\in\{e,p\},
\end{equation}
where $g=G\,M_{\odot}/R_{\odot}^{2}$ is the gravitational acceleration at the solar surface, \(G\) is the gravitational constant, \(M_{\odot}\) and \(R_{\odot}\) are the solar mass and radius, respectively, while \(m_e\) and \(m_p\) denote the electron and proton masses.

The plasma is also subjected to the equilibrium
Pannekoek--Rosseland electric field generated by stellar interior \citep{Pannekoek_1922,Rosseland_1924,Neslusan2001-rp,belmont2013collisionless},

\begin{equation}
\label{Pannekoekfield}
\mathbf{E}_{\mathrm{PR}}
=
-\frac{m_p-m_e}{2e} \mathbf{g}.
\end{equation}
The combined action of gravity and the Pannekoek--Rosseland electric field produces the same effective gravitational force on both species,
\begin{equation}
\mathbf{F}_{g,\alpha}
+
e_{\alpha}\mathbf{E}_{\mathrm{PR}}
=
m\,\mathbf{g},
\qquad
m
=
\frac{m_e+m_p}{2}.
\end{equation}
where \(e_{\alpha}=e\) for protons and \(e_{\alpha}=-e\) for electrons. In addition, particles could experience a supplementary electrostatic field \(E_s(z)\) generated by the electron-proton plasma of the atmosphere. In the present model, however, this supplementary field is expected a priori to vanish. Indeed, as we will show later when introducing the boundary conditions, the same heating is assumed for electrons and protons, while the Pannekoek--Rosseland electric field ensures that the two species experience the same external force \(m\,\mathbf{g}\) and therefore have the same scale height. No supplementary electrostatic field is thus expected to be required to maintain charge neutrality. This expectation will be demonstrated rigorously below, where we show that the self-consistent solution is \(E_s(z)=0\). Under these assumptions, the total force acting on the specy \(\alpha\) is
\begin{equation}
\label{totalforce}
\mathbf{F}_{\alpha}
=
m\mathbf{g}
+
e_{\alpha}
\left(
\mathbf{\Es}(z)
+
\frac{\mathbf{v}}{c}\times\mathbf{B}(z)
\right),
\end{equation}
$c$ is the speed of light. The particle dynamics is described along the axis of a vertical and expanding magnetic flux tube, with a field strength \(B(z)\) where \(z\) is the spatial vertical coordinate. \(B(z)\) is assumed to be a monotonically decreasing function of \(z\).

Indeed, magnetic flux tubes are generally expected to expand with height in stellar atmospheres \citep{Gabriel1976,Dowdy1986,Solanki_1991,Mandrini_2000,Judge_2021}. The cross-sectional area \(A(z)\) therefore increases with height so as to
conserve the magnetic flux,
\begin{equation}\label{magneticflux}
B(z)\,A(z)=\Phi_B=\mathrm{const}.
\end{equation}

\subsection{Collisionless dynamics} \label{sec:Collisionless_dynamics}

Since the plasma is assumed to be collisionless, the distribution function \(f_{\alpha}\) of each species satisfies the Vlasov equation
\begin{equation}
    \frac{\partial f_{\alpha}}{\partial t}
    +
    \mathbf{v}\cdot\nabla f_{\alpha}
    +
    \frac{\mathbf{F}_{\alpha}}{m_{\alpha}}
    \cdot
    \nabla_{\mathbf{v}}f_{\alpha}
    =
    0.
\end{equation}
The Vlasov equation describes an incompressible flow in single-particle phase space. Consequently, the distribution function is conserved along particle trajectories according to Liouville's theorem. The dynamics is therefore determined by the constants of motion.

The total particle energy is
\begin{equation}
\label{hamiltonian}
    \mathcal{H}_\alpha
    =
    \frac{1}{2}m_\alpha v^2
    +
    V_{\alpha}(z),
\end{equation}
where 
\begin{equation}
\label{def_V}
V_{\alpha}(z) = m\,g\,z + e_{\alpha} \phis (z) ,  
\end{equation}
and $\phis (z)$ is the electrostatic potential associated to $\Es$ with the boundary condition $\phis (0)=0$.

The velocity components parallel and perpendicular to the magnetic field are
\begin{equation}
    v_\parallel=v\cos\theta,
    \qquad
    v_\perp=v\sin\theta,
\end{equation}
where \(\theta\) is the pitch angle, namely the angle between the particle velocity and the local magnetic-field direction.

Within the guiding-centre approximation, the magnetic moment \citep{fitzpatrick2014plasma}
\begin{equation}
\label{magneticmoment}
    \mu_\alpha
    =
    \frac{m_\alpha v_\perp^2}{2B(z)}
    =
    \frac{m_\alpha v^2\sin^2\theta}{2B(z)}
\end{equation}
is an adiabatic invariant.

\subsection{Accessible velocity-space domain}  \label{sec:Accessible}

The populated region of velocity space is determined by requiring the parallel velocity to remain real. Starting from the conservation of total energy and magnetic moment, one finds
\begin{equation}
\sin^2\theta
=
\frac{B(z)}{B(0)}
\left(
1+
\frac{2\,V_\alpha(z)}{m_\alpha v^2}
\right)
\sin^2\theta_0 ,
\end{equation}
where \(0\le\sin^2\theta_0\le1\). Therefore,
\begin{equation}
\sin^2\theta
\le
\frac{B(z)}{B(0)}
\left(
1+
\frac{2\,V_\alpha(z)}{m_\alpha v^2}
\right).
\end{equation}
Using $\cos^2\theta = 1-\sin^2\theta$,
one immediately obtains
\begin{equation} \label{costheta_condition}
\cos^2\theta
\ge
1-
\frac{B(z)}{B(0)}
\left(
1+
\frac{2\,V_\alpha(z)}{m_\alpha v^2}
\right).
\end{equation}

The limiting trajectory separating particles that can reach height \(z\) from those that are mirrored corresponds to equality. In order that all $\theta$ values are possible the right hand side of Eq.~\eqref{costheta_condition} should vanish or be negative implying
\begin{equation} \label{def_vB}
v_{B,\alpha}^2(z)
=
\frac{B(z)}{B(0)-B(z)}
\frac{2\,V_\alpha(z)}{m_\alpha} \geq v^2.
\end{equation}
Introducing $v_{B,\alpha}$ in Eq.~\eqref{costheta_condition}, it rewrites as
\begin{equation}
\cos^2\theta
\ge
\left( 1-\frac{B(z)}{B(0)}
\right)
\left(
1-
\frac{v_{B,\alpha}^2(z)}{v^2}
\right).
\end{equation}
Defining
\begin{equation} \label{def_C(z)}
\mathcal{C}(z)
=
\sqrt{1-\frac{B(z)}{B(0)}},
\end{equation}
we get
\begin{equation}  \label{def_cos_theta_m}
\cos\theta
\ge \cos\theta_m
=
\mathcal{C}(z)
\sqrt{
1-
\frac{v_{B,\alpha}^2(z)}{v^2}
}.
\end{equation}
For $v<v_{B,\alpha}(z)$ the quantity under the square root becomes negative. In this case, all pitch angles remain connected to the lower boundary, and the entire velocity space is populated.

Conversely, for $v>v_{B,\alpha}(z)$ only particles with $|\theta| \le \theta_m $ can reach the height \(z\), whereas particles which would have larger pitch angles are reflected before reaching that altitude. Consequently, part of the velocity space cannot be populated by particles originating from the lower boundary, giving rise to the loss-cone region.

The distinction between the different regions of velocity space arising from the combined conservation of energy and magnetic moment, which defines the loss-cone structure, is commonly employed in exospheric models of the solar wind \cite{Jockers1970,Lemaire1971,Pierrard1996,Maksimovic1997,Lamy2003,Zouganelis2004}. In particular, in \cite{Zouganelis2004}, the different accessible regions and the subsequent calculation of the moments of the distribution function were formulated directly in the \((\mathcal{H},\mu)\) space rather than in the \((v_{\parallel},v_{\perp})\) space adopted here.

Although this procedure for determining the accessible region of phase space is quite general, exospheric models of the solar wind must additionally account for the open magnetic geometry and the presence of escaping particles, which give rise to the solar-wind outflow. The presence of these escaping particles introduces an additional constraint in velocity space and results in a finite bulk velocity. This outflow also modifies the self-consistent electrostatic field generated by the plasma \cite{Lemaire1971}. In the notation adopted here, this additional contribution would be associated with \(E_s\), and its presence would modify the results obtained for an open magnetic geometry. In such a configuration, \(E_s\) must be determined by simultaneously imposing the conditions of zero net charge density and zero electric current. The resulting algebraic equations generally need to be solved numerically.

In the present model, however, the magnetic configuration is closed and no stellar wind develops, as discussed at the beginning of Sect.~\ref{sec:Basic_equations}. Therefore, this additional effect is absent, and no numerical solution of the zero-current condition is required. This simplification allows the self-consistent electrostatic problem, and consequently the subsequent calculations, to be treated fully analytically.

\section{Single-temperature solution}
\label{sec:single-temperature}

\subsection{Boundary conditions}
\label{sec:Boundary_conditions}

At the base of the plasma slab (\(z=0\)), the system is assumed to be in
contact with a thermal reservoir. The reservoir injects particles into the loop with a
Maxwellian distribution,
\begin{equation}
\label{boundary}
\begin{aligned}
f_{T,\alpha}(\mathbf v)
={}&
\nb
\left(
\frac{m_\alpha}{2\pi k_B \Tb}
\right)^{3/2}
\\
&\times
\exp\!\left(
-\frac{m_\alpha |\mathbf v|^2}{2k_B \Tb}
\right),
\qquad
v_\parallel>0.
\end{aligned}
\end{equation}

Here \(v_\parallel\) is the velocity component along the magnetic field,
directed away from the lower boundary, and \(\nb\) and \(\Tb\) denote the
density and temperature at the base of the model. 

\subsection{Analytical solution} \label{sec:Analytical_solution}

When the temperature $\Tb$ of the Maxwellian distribution injected at the lower boundary is fixed, the distribution function at height \(z\) is obtained
by Liouville mapping along the trajectories connected to the lower boundary. The mapped distribution is therefore
\begin{equation}
\label{def_f}
\begin{aligned}
f_{T,\alpha}(z,\mathbf v)
={}&
\nb
\left(
\frac{m_\alpha}{2\pi k_B \Tb}
\right)^{3/2}
\\
&\times
\exp\!\left(
-\frac{\mathcal{H}_{\alpha}}{k_B \Tb}
\right)
\chi(z,\mathbf v).
\end{aligned}
\end{equation}
where \(\chi(z,\mathbf v)\) is equal to unity for trajectories connected to the lower boundary and zero otherwise.  Equivalently,
\begin{equation}
\label{velocityfilter}
\chi(z,\mathbf v)=
\begin{cases}
1,
& v \le v_{B,\alpha}(z),
\\[3pt]
1,
& \begin{aligned}[t]
v>v_{B,\alpha}(z),\\
|\theta|\le\theta_m(z,v),
\end{aligned}
\\[3pt]
0,
& \begin{aligned}[t]
v>v_{B,\alpha}(z),\\
|\theta|>\theta_m(z,v).
\end{aligned}
\end{cases}
\end{equation}

Thus, the Maxwellian factor in Eq.~\eqref{def_f} should not be interpreted as a full local Maxwellian at height \(z\). It is the Liouville-mapped value of the
distribution on the portion of phase space connected to the lower boundary. In the presence of magnetic expansion, particles with sufficiently large pitch angles are reflected before reaching the position \(z\), producing a depletion of the distribution function within the loss-cone region.

The detailed derivation of the moments of the distribution function is presented in Appendix \ref{appendix:moments}. The following quantity appears in the analytical integrations:
\begin{equation} \label{def_a(z,T)}
\as_{\alpha}(z)
=
-\frac{V_{\alpha}(z)}{k_B \Tb}
\left(
1-\frac{1}{\mathcal{C}^2(z)}
\right),
\end{equation}
with $\mathcal{C}(z)$ defined by Eq.~\eqref{def_C(z)}. The quantity \(a_{\alpha}(z)\) measures the relative importance of the potential energy with respect to the thermal energy, modulated by the magnetic expansion factor.  Below we report the final expressions of density and temperatures and discuss their physical interpretation.

\subsubsection{Number density and quasi-neutrality}

Integrating the distribution function over the accessible velocity domain yields the density
\begin{equation}
\label{densitysingletemp}
n_{\alpha}(z)
=
\nb \,
e^{-V_{\alpha}(z)/(k_B \Tb)}
\left[
1-\mathcal{C}(z)e^{-a_{\alpha}(z)}
\right].
\end{equation}
The first exponential factor corresponds to the standard barometric stratification, whereas the second accounts for the removal of inaccessible regions of velocity space. In the absence of magnetic expansion, namely in the limit $\mathcal{C}(z)\rightarrow 0 $ the standard barometric density profile is recovered.

We determine the supplementary electrostatic potential $\phis$ by imposing local charge
neutrality,
\begin{equation}
\label{eq:neutrality}
n_e(z)=n_p(z).
\end{equation}
Introducing the auxiliary quantities
\[
X_{\pm}(z,\phis)
=
\frac{\pm e\phis+m g z}{k_B\Tb},
\qquad
D(z)
=
1-\frac{1}{\mathcal{C}^2(z)},
\]
the charge density difference can be written as
\begin{equation}
\label{F}
\begin{aligned}
F(\phis)
={}& n_p(z)-n_e(z)
\\
={}&
e^{-X_{+}(z,\phis)}
\left[
1-\mathcal{C}(z)
e^{-X_{+}(z,\phis)D(z)}
\right]
\\
&-
e^{-X_{-}(z,\phis)}
\left[
1-\mathcal{C}(z)
e^{-X_{-}(z,\phis)D(z)}
\right].
\end{aligned}
\end{equation}

The neutrality condition is therefore equivalent to
\begin{equation}
\label{zerosphi}
F(\phis)=0.
\end{equation}
which is satisfied by
\begin{equation}\label{zeroelectricpotential}
\phis(z)=0 .
\end{equation}

To establish uniqueness, it is sufficient to study the monotonicity of
\(F\). Differentiating Eq.~\eqref{F}, one finds
\begin{equation}
\frac{dF}{d\phis}<0
\end{equation}
for every value of \(\phis\) (see Appendix \ref{appendix:derivative}). Hence \(F(\phis)\) is strictly decreasing and can possess at most one
zero. Since \(F(0)=0\), Eq.~\eqref{zerosphi} has the unique solution given by Eq. \eqref{zeroelectricpotential}. Therefore, local charge neutrality uniquely determines that no extra electric field is present in the single-temperature collisionless solution. The resulting total potential is equal for both species and is given by
\begin{equation}  \label{totalpotential}
V_\alpha(z)=m\,g\,z.
\end{equation}

The loss-cone parameter, defined by Eq.~\eqref{def_a(z,T)}, therefore becomes
\begin{equation}\label{apanne}
\az
=
\frac{z}{\Hg}
\left(
\frac{1}{\mathcal{C}^2(z)}-1
\right),
\end{equation}
where 
\begin{equation}\label{def_H}
\Hg = \frac{k_B T_b}{mg}
\end{equation}
is the gravitational scale height.
Since the magnetic field decreases monotonically with height, one has
\begin{equation} \label{propertiesc}
0 \le \mathcal{C}(z) \le 1 \,, \mbox{then } \az \ge 0.
\end{equation}
The density, Eq. \eqref{densitysingletemp}, is then given by
\begin{equation}  \label{densitypanne}
n(z)
=
\nb\,
e^{-\frac{z}{\Hg}}
\left[
1-\mathcal{C}(z)e^{-\az}
\right].
\end{equation}

\subsubsection{Temperatures}

The conservation of energy and magnetic moment determines the subset of particle trajectories that remain magnetically connected to the lower boundary. The collisionless Liouville mapping therefore populates only a restricted region of velocity space. The resulting truncation of the distribution function (loss cone) produces an anisotropic velocity distribution, causing the parallel and perpendicular velocity moments to evolve differently along the flux tube.

The parallel temperature is
\begin{equation}
\label{paralleltemperaturepanne}
T_{\parallel}(z)
=
\Tb
\frac{
1-\mathcal{C}^3(z)\,e^{-\az}
}{
1-\mathcal{C}(z)\,e^{-\az}
},
\end{equation}
whereas the perpendicular temperature is
\begin{equation}
\label{perpendicultemperaturepanne}
\begin{aligned}
T_{\perp}(z)
={}&
\Tb
\Biggl[
1
-
\mathcal{C}(z)
\left(
\frac{3}{2}+\az
\right)
e^{-\az}
\\
&\qquad
+
\frac{1}{2}
\mathcal{C}^3(z)e^{-\az}
\Biggr]
\\
&\times
\left[
1-\mathcal{C}(z)e^{-\az}
\right]^{-1}.
\end{aligned}
\end{equation}

The perpendicular temperature $T_{\perp} (z)$ is more strongly affected than $T_{\parallel}(z)$ because the
loss-cone structure selectively removes particles with large pitch angles, corresponding to large perpendicular kinetic energies. As a consequence of Eq.~\eqref{propertiesc}, $\mathcal{C}^3(z) \le \mathcal{C}(z) $, then
\begin{equation}\label{temperatureimbalance}
T_{\perp} (z)
\le
T_{\parallel}(z),
\end{equation}
and the plasma develops a pressure anisotropy.

The total temperature is defined as
\begin{equation}\label{totaltemperaturedef}
T(z)
=
\frac{
T_{\parallel}(z)
+
2\,T_{\perp}(z)
}{3},
\end{equation}
which yields  
\begin{equation}
\label{totaltemperaturepanne}
T(z)
=
\Tb \left[ 1 - \frac{2}{3}
\frac{
\mathcal{C}(z)\,\az\,e^{-\az}
}{
1-\mathcal{C}(z)\,e^{-\az}
}
\right],
\end{equation}
Using Eq.~\eqref{propertiesc}, this expression shows that the combined action of gravity and magnetic field produces an effective cooling, $T(z) \le \Tb$, of the particle population with height. Physically, this cooling arises because the high-energy portion of velocity space becomes progressively depleted as particles with unfavorable pitch angles are reflected before reaching the upper regions of the flux tube.

\subsection{General properties of the single-temperature solution}
\label{sec:General_properties}

Using Eq.~\eqref{propertiesc} it follows that
\begin{equation}\label{propertiesac}
0 \le
1-\mathcal{C}(z)e^{-\az}
\le1.
\end{equation}
Using the inequality above together with
Eq.~\eqref{densitypanne}, we obtain
\begin{equation}
n(z)
\le
\nb\,
e^{-\frac{z}{\Hg}},
\end{equation}
showing that the combined effect of conservation of energy and magnetic moment always reduces the density with respect to the corresponding unmagnetized isothermal atmosphere. This reduction originates from the loss-cone depletion of the accessible velocity space, which acts in addition to gravitational filtering.

Using the expression for the parallel temperature given by
Eq.~\eqref{paralleltemperaturepanne}, together with the inequality
\(\mathcal{C}^3(z) \le \mathcal{C}(z)\), it immediately follows that
\begin{equation}
\label{parallelequality}
T_{\parallel}(z)\ge \Tb.
\end{equation}
The opposite result is obtained for the perpendicular temperature,
\begin{equation} \label{constraintTperp}
T_{\perp}(z) \le \Tb.
\end{equation}
since the numerator and denominator of Eq.~\eqref{perpendicultemperaturepanne} satisfy
\begin{equation}
1
-
\mathcal{C}
\left(
\frac32+\as
\right)
e^{-\as}
+
\frac12 \mathcal{C}^3 e^{-\as}
\le
1
-
\mathcal{C}\, e^{-\as},
\end{equation}
which can be rewritten as
\begin{equation}
\frac12 \mathcal{C}^2
\le
\frac12+\as,
\end{equation}
and, due to Eq.~\eqref{propertiesc},
Eq. \eqref{constraintTperp} is always satisfied. Therefore, the perpendicular temperature is necessarily smaller than the boundary temperature.

Furthermore, using Eq.~\eqref{totaltemperaturedef},
\begin{equation}
T(z)-T_{\perp}(z)
=
\frac{
T_{\parallel}(z)-T_{\perp}(z)
}{3},
\end{equation}
and recalling that $T_{\perp} \le T_{\parallel}$, Eq.~\eqref{temperatureimbalance}, it follows that 
\begin{equation}
\label{perpendicularequality}
T(z) \ge T_{\perp}(z).
\end{equation}

Combining all previous inequalities, one obtains the ordering
\begin{equation}
\label{orderingtemperatures}
T_{\perp}(z)
\le
T(z)
\le
\Tb
\le
T_{\parallel}(z).
\end{equation}

This hierarchy provides a simple physical interpretation of the combined effect of energy and magnetic-moment conservation. Together, these conservation laws progressively restrict the region of velocity space accessible to particles reaching a given height, producing the loss-cone anisotropy. As a consequence, the local velocity distribution becomes increasingly elongated along the magnetic field. The perpendicular second velocity moment is therefore reduced, leading to a decrease of both the perpendicular and total temperatures, whereas the parallel second moment is enhanced, yielding a parallel temperature above its boundary value.

\subsection{Spatial limits of the single-temperature solution}
\label{sec:Spatial_limits}

At the lower boundary,
\begin{equation}
z\rightarrow0,
\qquad
\mathcal{C}(z)\rightarrow0,
\end{equation}
the loss-cone correction vanishes. Consequently,
\begin{equation}
\label{densitybase}
n(z)\rightarrow\nb,
\end{equation}
while
\begin{equation}
\label{temperaturesbase}
T_{\parallel}(z)\rightarrow\Tb,
\qquad
T_{\perp}(z)\rightarrow\Tb,
\qquad
T(z)\rightarrow\Tb.
\end{equation}
The velocity distribution therefore remains isotropic in the vicinity
of the lower boundary.

We now turn to the large-height limit. We assume that
\begin{equation}
\label{asymptoticB}
B(z)\rightarrow0,
\qquad
\mathcal{C}(z)\rightarrow1,
\end{equation}
and that the magnetic-field strength decreases faster than \(z^{-1}\),
namely,
\begin{equation}
\label{magneticlimit}
z\,B(z)\rightarrow0,
\end{equation}

Using Eq.~\eqref{def_C(z)} and denoting \(B(0)=B_b\), the loss-cone parameter becomes
\begin{equation}
\label{asymptotica}
\az
=
\frac{z}{\Hg}
\frac{1-\mathcal{C}^2(z)}
{\mathcal{C}^2(z)}
\simeq
\frac{z}{\Hg}
\frac{B(z)}{\Bb}.
\end{equation}
Equation~\eqref{magneticlimit} therefore guarantees that
\(\az\rightarrow 0\) at large heights.

\subsubsection{Density}

The density is given by Eq.~\eqref{densitypanne}. At large heights,
combining Eqs.~\eqref{asymptoticB} and \eqref{asymptotica} yields
\begin{equation}
\label{asymptoticaC}
1-\mathcal{C}(z)e^{-\az}
\simeq
\az.
\end{equation}
Substituting this result into Eq.~\eqref{densitypanne}, we obtain the
asymptotic density profile,
\begin{equation}
\label{density_large_z}
n(z)
\simeq
\nb
e^{-z/\Hg}
\az
\simeq
\nb
e^{-z/\Hg}
\frac{z}{\Hg}
\frac{B(z)}{\Bb}.
\end{equation}

In particular, if the magnetic-field strength decreases asymptotically
as
\begin{equation}
\label{magneticpowerlaw}
B(z)\propto z^{-p},
\qquad
p>1,
\end{equation}
then
\begin{equation}
n(z)
\propto
e^{-z/\Hg}
z^{1-p}.
\end{equation}
Thus, the combined conservation of energy and magnetic moment produces an additional algebraic depletion of the density, proportional to \(zB(z)\), on top of the ordinary barometric stratification.

\subsubsection{Parallel temperature}

The parallel temperature is given by
Eq.~\eqref{paralleltemperaturepanne}. Expanding the exponential for
\(\az\ll1\),
\begin{equation}
e^{-\az}
\simeq
1-\az,
\end{equation}
and using Eq.~\eqref{asymptotica}, we obtain
\begin{equation}
\label{paralleltemperature_large_z}
T_{\parallel}(z)
\simeq
\Tb
\frac{
1+\mathcal{C}+\mathcal{C}^2
+
\mathcal{C}(1+\mathcal{C})z/\Hg
}{
1+(1+\mathcal{C})z/(\Hg\mathcal{C})
}.
\end{equation}

Since the terms proportional to \(z/\Hg\) dominate asymptotically,
\begin{equation}
\label{paralleltemperature_large_z_gravity}
\lim_{z\rightarrow\infty}
T_{\parallel}(z)
=
\Tb.
\end{equation}
The parallel temperature therefore approaches the boundary value both at the lower boundary and in the asymptotic limit. If \(T_{\parallel}(z)\) exceeds the boundary temperature at intermediate altitudes, this enhancement must eventually disappear as \(z\rightarrow\infty\). The detailed evolution of
\(T_{\parallel}(z)\), including the possible existence and number of extrema, depends on the specific magnetic-field profile and cannot be deduced from the asymptotic analysis alone.

The physical origin of this behaviour can nevertheless be understood as follows. Close to the lower boundary, the loss cone is still narrow, and only a small fraction of particles is removed. As the magnetic
field expands, particles with large pitch angles are preferentially excluded, leading to a selective depletion of the perpendicular component of the velocity distribution. The surviving population
therefore becomes increasingly elongated along the magnetic field, enhancing the parallel second velocity moment and, consequently, the parallel temperature.

At larger heights, the loss cone occupies an increasingly large fraction of velocity space, progressively depleting the particle population connected to the lower boundary. Although the remaining
particles are still preferentially field aligned, both the density and the parallel second velocity moment acquire the same leading dependence. Their ratio therefore gradually returns to the boundary value, causing the parallel temperature to asymptotically
approach \(\Tb\).

\subsubsection{Perpendicular temperature}

The perpendicular temperature is given by
Eq.~\eqref{perpendicultemperaturepanne}. At large heights, the denominator behaves according to Eq.~\eqref{asymptoticaC}. Next, the exponential must be
expanded to second order in the numerator, since the constant and linear terms cancel. One obtains
\begin{align}
&
1
-
\mathcal{C}(z)
\left(
\frac{3}{2}+\az
\right)
e^{-\az}
+
\frac{1}{2}
\mathcal{C}^3(z)e^{-\az}
\nonumber\\
&\qquad
\simeq
\frac{\az^2}{2}.
\end{align}
Therefore,
\begin{equation}
\label{perpendicultemperature_large_z}
T_{\perp}(z)
\simeq
\frac{\Tb\,\az}{2}
\simeq
\frac{\Tb}{2}
\frac{z}{\Hg}
\frac{B(z)}{\Bb}.
\end{equation}

Since \(zB(z)\rightarrow0\),
\begin{equation}
\label{perpendicultemperature_large_z_limit}
\lim_{z\rightarrow\infty}
T_{\perp}(z)
=
0.
\end{equation}
Furthermore, if the magnetic field satisfies the asymptotic power law given by Eq.~\eqref{magneticpowerlaw},
\begin{equation}
T_{\perp}(z)
\propto
z^{1-p}.
\end{equation}

Thus, the perpendicular temperature exhibits an algebraic decay at large heights, with an asymptotic behaviour directly controlled by the magnetic-field profile. This prediction will be confirmed numerically
in the next subsection. The underlying physical mechanism can be understood as follows.

The combined conservation of energy and magnetic moment progressively restricts the region of velocity space accessible to particles reaching a given height. As the magnetic field expands, the allowed range of pitch angles becomes increasingly narrow, and the local
velocity distribution is progressively focused along the magnetic field. Consequently, the perpendicular second velocity moment is continuously reduced. Since the perpendicular temperature is directly proportional to this second moment, it is progressively depleted as
the plasma propagates through the expanding magnetic flux tube.

At sufficiently large heights, the loss cone occupies most of the accessible velocity space, and only particles with nearly field-aligned velocities remain magnetically connected to the lower boundary. As a consequence, the perpendicular temperature progressively vanishes, approaching zero asymptotically.

\subsubsection{Total temperature}

The total temperature is given by Eq.~\eqref{totaltemperaturepanne}. Combining Eq.~\eqref{asymptoticaC} with
\begin{equation}
\mathcal{C}(z)e^{-\az} \rightarrow 1, \qquad z \rightarrow +\infty,
\end{equation}
immediately yields
\begin{equation}
\label{totaltemperature_large_z_limit}
\lim_{z\rightarrow\infty}
T(z)
=
\frac{\Tb}{3}.
\end{equation}

The leading correction to this asymptotic limit can also be obtained explicitly. Using
\begin{equation}
1-\mathcal{C}(z)
\simeq
\frac{1}{2}
\frac{B(z)}{\Bb},
\end{equation}
one finds
\begin{equation}
1-\mathcal{C}e^{-\az}
\simeq
\az
+
(1-\mathcal{C})
-
\frac{\az^2}{2},
\end{equation}
and
\begin{equation}
\mathcal{C}\,\az\, e^{-\az}
\simeq
\az
-
\az^2
-
\az(1-\mathcal{C}).
\end{equation}
Therefore,
\begin{equation}
\frac{
\mathcal{C}\,\az\, e^{-\az}
}{
1-\mathcal{C}e^{-\az}
}
\simeq
1
-
\frac{\az}{2}
-
\frac{1-\mathcal{C}}{\az},
\end{equation}
which, substituted into Eq.~\eqref{totaltemperaturepanne}, gives
\begin{equation}
\label{totaltemperature_large_z}
T(z)
\simeq
\frac{\Tb}{3}
\left[
1
+
\az
+
2\frac{1-\mathcal{C}}{\az}
\right].
\end{equation}

Finally, using Eqs.~\eqref{asymptotica} together with
\begin{equation}
1-\mathcal{C}(z)
\simeq
\frac{1}{2}
\frac{B(z)}{\Bb},
\end{equation}
the general asymptotic behaviour becomes
\begin{equation}
\label{totaltemperature_large_z_B}
T(z)
\simeq
\frac{\Tb}{3}
\left[
1
+
\frac{z}{\Hg}
\frac{B(z)}{\Bb}
+
\frac{\Hg}{z}
\right].
\end{equation}
Both correction terms vanish under the condition
\eqref{magneticlimit}, consistently with the asymptotic limit given by Eq.~\eqref{totaltemperature_large_z_limit}. Their relative importance, however, depends on the asymptotic behaviour of the magnetic field. In particular, if the magnetic field satisfies the power-law behaviour
given by Eq.~\eqref{magneticpowerlaw}, the magnetic correction scales as \(z^{1-p}\), whereas the second correction always decreases as \(z^{-1}\).

The physical interpretation of this result follows directly from the different asymptotic behaviours of the parallel and perpendicular temperature components. At large heights, the perpendicular temperature vanishes asymptotically according to Eq.~\eqref{perpendicultemperature_large_z_limit}, whereas the parallel temperature approaches the finite limit given by Eq.~\eqref{paralleltemperature_large_z_gravity}. Consequently, only one of the three velocity degrees of freedom retains a finite temperature, while the other two become progressively depleted. Using the definition
of the total temperature, Eq.~\eqref{totaltemperaturedef}, one therefore recovers the asymptotic limit \(T(z)\rightarrow\Tb/3\).

The asymptotic velocity distribution is therefore strongly anisotropic, being increasingly concentrated along the magnetic-field direction. Nevertheless, the asymptotic analysis alone does not establish whether
\(T(z)\) is monotonic for an arbitrary magnetic-field profile.

\subsubsection{Limit without gravity}

In the absence of gravity, the loss-cone parameter \(a(z)\) vanishes.
The density, Eq.~\eqref{densitypanne}, reduces to
\begin{equation}
\label{density_no_gravity}
n(z,g=0)
=
\nb
\left[
1-\mathcal{C}(z)
\right].
\end{equation}

Since the magnetic-field strength decreases monotonically with height, \(\mathcal{C}(z)\) increases monotonically from \(0\) to \(1\), and the density decreases monotonically. At large heights,
\begin{equation}
n(z,g=0)
\simeq
\frac{\nb}{2}
\frac{B(z)}{\Bb},
\end{equation}
showing that, in the absence of gravity, conservation of magnetic moment alone produces an algebraic density depletion controlled solely by the magnetic-field expansion.

The parallel and perpendicular temperatures become
\begin{equation}
\label{paralleltemperaturepanne_no_gravity}
T_{\parallel}(z,g=0)
=
\Tb
\left[
1+\mathcal{C}(z)+\mathcal{C}^2(z)
\right],
\end{equation}
and
\begin{equation}
\label{perpendicultemperaturepanne_no_gravity}
T_{\perp}(z,g=0)
=
\Tb
\left[
1-\frac{\mathcal{C}(z)}{2}
-\frac{\mathcal{C}^2(z)}{2}
\right].
\end{equation}

Since \(\mathcal{C}(z)\) increases monotonically with height, \(T_{\parallel}(z)\) increases monotonically, whereas \(T_{\perp}(z)\) decreases monotonically. Their asymptotic limits are
\begin{equation} \label{asymptoticTgzero}
\lim_{z\rightarrow\infty}
T_{\parallel}(z,g=0)
=
3\,\Tb,
\qquad
\lim_{z\rightarrow\infty}
T_{\perp}(z,g=0)
=
0.
\end{equation}

The monotonic increase of the parallel temperature contrasts with its behaviour in the presence of gravity. Without gravitational stratification, there is no progressive depletion of the particle population with height. Consequently, the preferential transfer of
kinetic energy from the perpendicular to the parallel degrees of freedom produced by magnetic moment conservation is never compensated, and the
parallel temperature increases continuously until reaching its asymptotic value \(3\,\Tb\).

It is worth emphasizing that this asymptotic limit is obtained by taking first the limit \(\Hg\rightarrow\infty\) in Eq.~\eqref{paralleltemperature_large_z}. The limits
\(\Hg\rightarrow\infty\) and \(z\rightarrow\infty\) therefore do not commute. Indeed, for any finite value of \(\Hg\), the ratio \(z/\Hg\) eventually becomes arbitrarily large as \(z\rightarrow\infty\), leading instead to the asymptotic limit
\begin{equation}
\lim_{z\rightarrow\infty}
T_{\parallel}(z)
=
\Tb.
\end{equation}

The total temperature remains exactly constant,
\begin{equation}
T(z,g=0)
=
\frac{
T_{\parallel}(z,g=0)
+
2T_{\perp}(z,g=0)
}{3}
=
\Tb.
\end{equation}

Thus, in the absence of gravity, conservation of magnetic moment merely redistributes kinetic energy between the parallel and perpendicular degrees of freedom, while leaving the total kinetic temperature
unchanged. This follows directly from the fact that the Lorentz force performs no work on charged particles.

\subsection{Atmosphere properties above a magnetic concentration} \label{sec:Magnetic_field}
\subsubsection{Magnetic field above a concentrated source}
In order to illustrate the effects of magnetic moment conservation, we adopt a specific model for the magnetic-field expansion. We consider a potential magnetic field above a concentrate source located at depth $z=-d$., The magnetic-field strength along the reference field line decreases with height according to
\begin{equation}\label{Bpotential}
B(z)
=
\Bb
\frac{d^2}{z^2+d^2},
\end{equation}
where \(\Bb\) is the magnetic-field strength at the lower boundary and \(d\) denotes the depth of the equivalent magnetic source below the
base of the model. This profile provides a simple analytical representation of the magnetic-field expansion in stellar atmospheres and has been widely adopted to model expanding coronal magnetic flux
tubes \citep[e.g.][]{Demoulin_1994}.

Using magnetic-flux conservation, Eq.~\eqref{magneticflux}, the cross-sectional area of the flux tube is found to evolve as
\begin{equation}
A(z)
\propto
\frac{1}{B(z)}
\propto
z^2+d^2,
\end{equation}
showing that the flux tube expands monotonically with height. With this choice, the magnetic expansion parameter becomes
\begin{equation}\label{cpotential}
\mathcal{C}(z)
=
\sqrt{
1-\frac{B(z)}{\Bb}
}
=
\frac{z}{\sqrt{z^2+d^2}}.
\end{equation}

At large heights (\(z\gg d\)), the magnetic-field strength behaves as
\begin{equation}
B(z)
=
\Bb
\frac{d^2}{z^2+d^2}
\simeq
\Bb\,
\frac{d^2}{z^2},
\end{equation}
which is a particular realization of the general class of magnetic fields considered in Section~\ref{sec:Spatial_limits}, satisfying
\(B(z)=\mathcal{O}(z^{-1})\). Substituting this asymptotic behaviour into the general expressions derived in Section~\ref{sec:Spatial_limits} yields
\begin{equation}
n(z)
\simeq
\nb\,
e^{-z/\Hg}
\frac{d^2}{\Hg\,z},
\end{equation}
and
\begin{equation}
T_{\perp}(z)
\simeq
\Tb\,
\frac{d^2}{2\Hg\,z},
\end{equation}
while the asymptotic limits for parallel temperature given by Eq.\eqref{paralleltemperature_large_z_gravity} and for total temperature given by Eq. \eqref{totaltemperature_large_z_limit} remain unchanged.

All the results presented below are obtained using this magnetic-field profile. Nevertheless, although the quantitative values depend on the specific form of \(B(z)\), the physical mechanisms discussed in this
work are generic consequences of any magnetic field whose strength decreases monotonically with height.

\begin{figure*}
\centering
\includegraphics[width=0.99\textwidth]{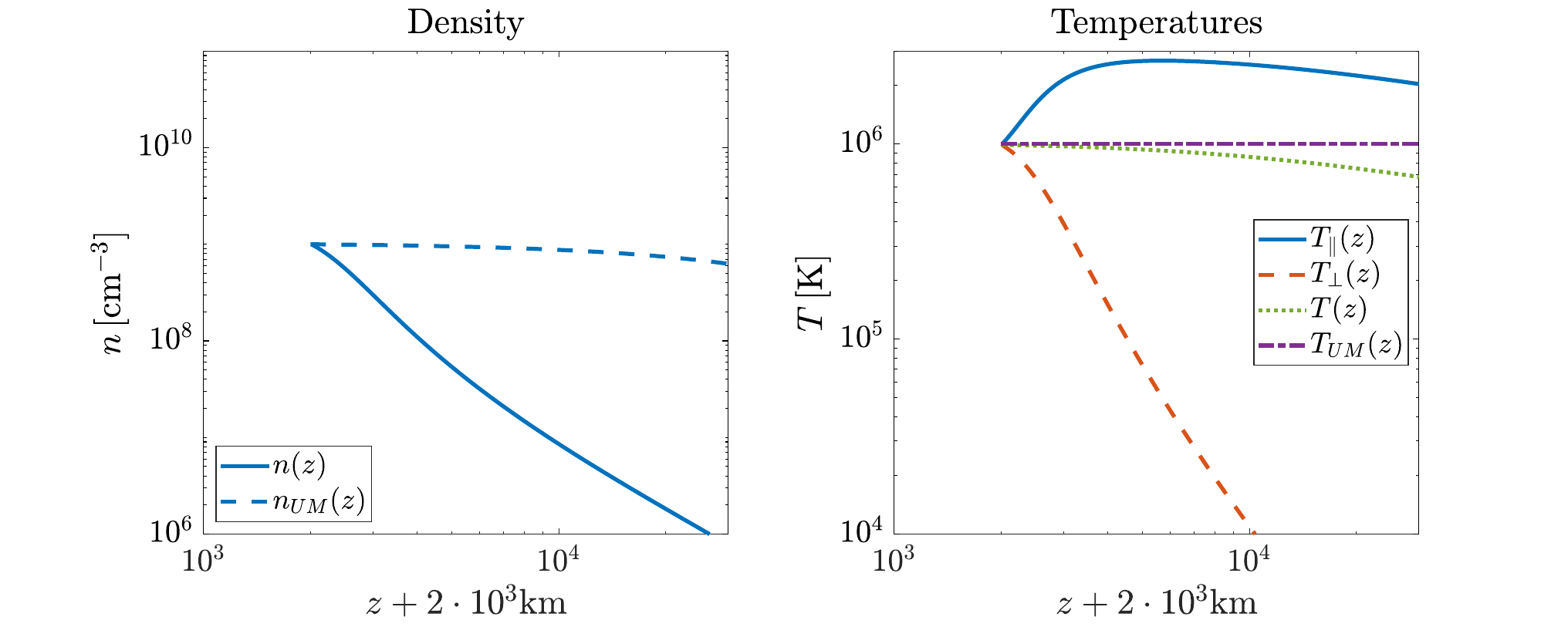}
\caption{Left panel: density profile \(n(z)\) computed from Eq.~\eqref{densitypanne} and compared with the corresponding unmagnetized profile \(n_{\rm UM}(z)\) given by Eq.~\eqref{densityunmagnetized}. Right panel: total, parallel, and perpendicular temperature profiles, \(T(z)\), \(T_{\parallel}(z)\), and \(T_{\perp}(z)\), computed from Eqs.~\eqref{totaltemperaturepanne}, \eqref{paralleltemperaturepanne}, and \eqref{perpendicultemperaturepanne}, respectively. The corresponding unmagnetized temperature profile \(T_{\rm UM}(z)\), computed from Eq.~\eqref{temperatureunmagnetized}, is also shown. The colour scheme, line styles, and legend conventions are indicated in the figure. The parameters are \(\Tb=10^{6}\,\mathrm{K}\), \(\nb=10^{9}\,\mathrm{cm^{-3}}\), \(d=10^{3}\,\mathrm{km}\).}
\label{fig1}
\end{figure*}

\begin{figure*}
\centering
\includegraphics[width=0.99\textwidth]{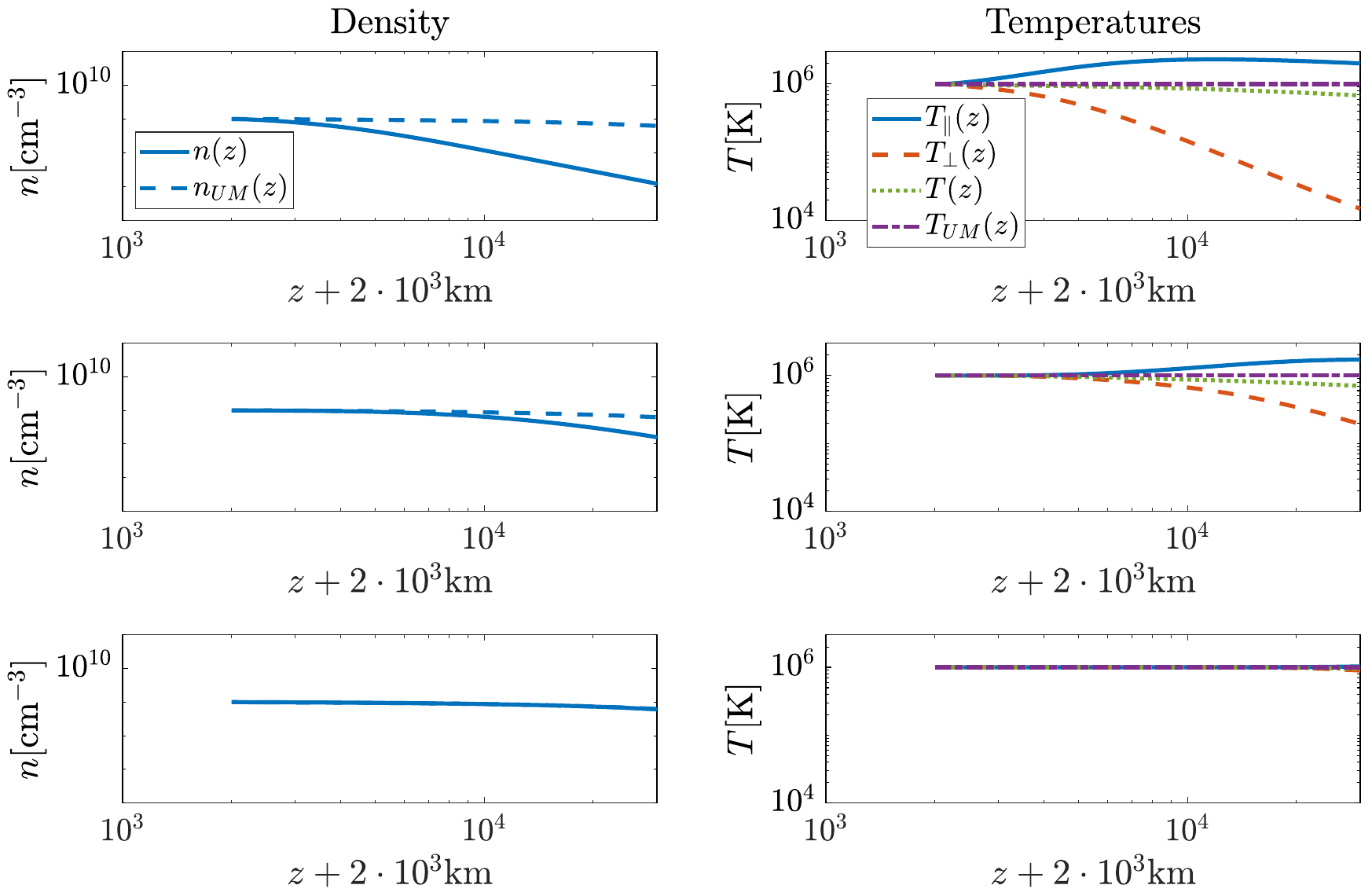}
\caption{
Comparison of the single-temperature solutions for different magnetic-field expansion lengths \(d\). The first row corresponds to \(d=4\times10^3\,\mathrm{km}\), the second row to
\(d=1.6\times10^4\,\mathrm{km}\), and the third row to
\(d=6.4\times10^4\,\mathrm{km}\). In each row, the left panel compares the magnetized and unmagnetized density profiles computed from Eqs.~\eqref{densitysingletemp} and \eqref{densityunmagnetized}, respectively. The right panel shows the corresponding parallel,
perpendicular, total, and unmagnetized temperature profiles computed from Eqs.~\eqref{paralleltemperaturepanne}, \eqref{perpendicultemperaturepanne}, \eqref{totaltemperaturedef}, and \eqref{temperatureunmagnetized}. The colour scheme, line styles, legend conventions, and all remaining parameters are identical to those adopted in Figure~\ref{fig1}.
}
\label{fig2}
\end{figure*}

\begin{figure*}
\centering
\includegraphics[width=0.99\textwidth]{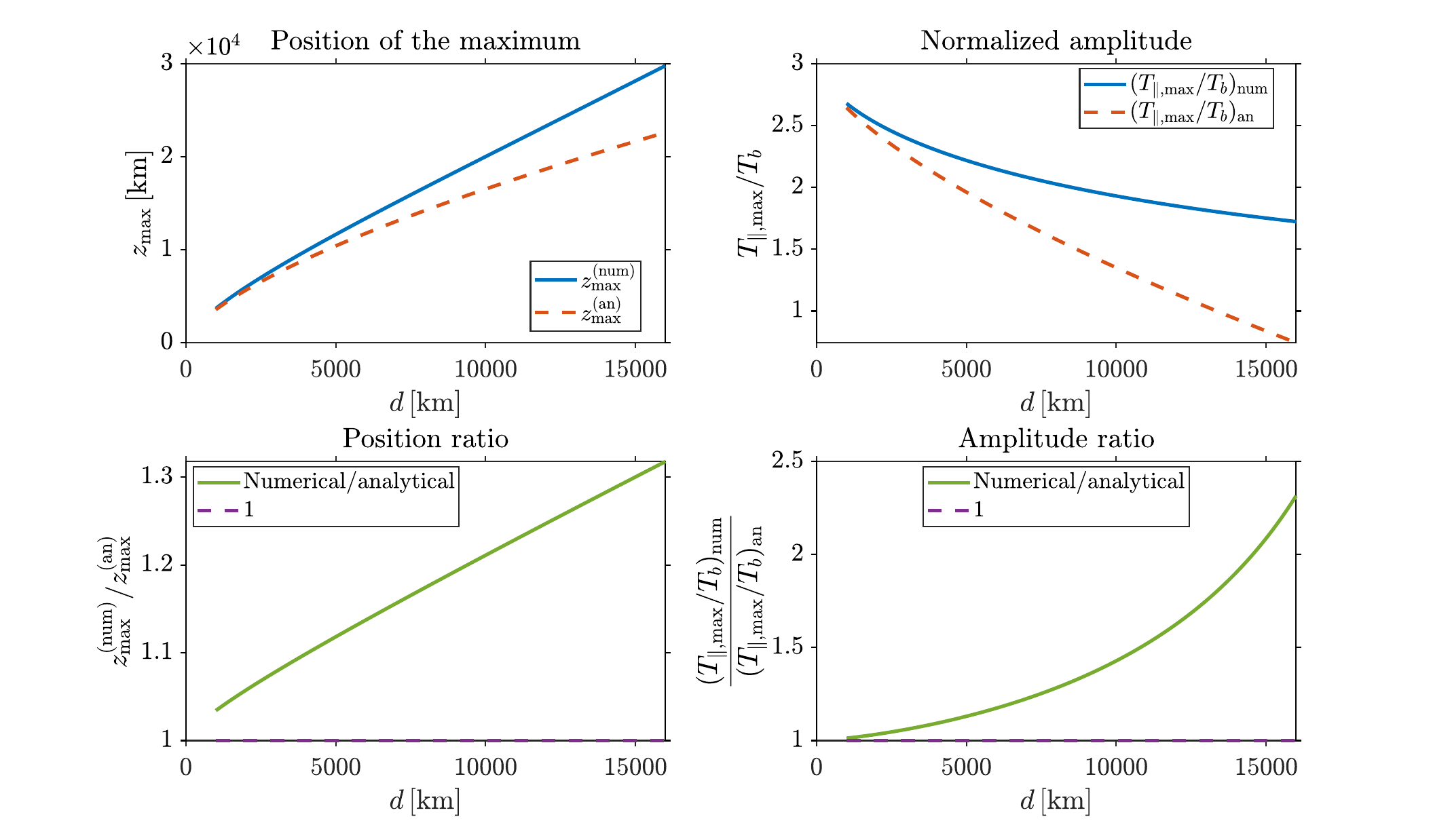}
\caption{Comparison between the analytical predictions and the numerical solution for the magnetic-field profile given by Eq.~\eqref{Bpotential}. Top left: location of the maximum parallel temperature as a function of the magnetic scale \(d\). The analytical prediction is given by Eq.~\eqref{eq:zmax_generalized_potential} (red dashed curve), while the numerical solution is obtained from Eq.~\eqref{numericalzmax} (blue continuos curve). Top right: corresponding maximum parallel temperature normalized to the basal temperature \(T_b\). The analytical prediction is given by Eq.~\eqref{eq:tparallelmax_generalized}(red dashed curve), while the numerical solution is obtained from Eq.~\eqref{numericalTmax}(blue continuos curve). Bottom panels: ratios between the numerical and analytical predictions for the location (left) and the amplitude (right) of the maximum.}
\label{fig3}
\end{figure*}

\subsubsection{Density and temperature profiles}\label{subsubsec:densitytempsingle}

We compare the density and temperature profiles obtained in the presence of magnetic expansion with the corresponding unmagnetized solutions. In the absence of magnetic expansion, the density reduces to the standard
Maxwell-Boltzmann barometric profile,
\begin{equation}
\label{densityunmagnetized}
n_{\rm UM}(z)
=
\nb
\exp\!\left(
-\frac{z}{\Hg}
\right),
\end{equation}
while the temperature remains spatially uniform,
\begin{equation}
\label{temperatureunmagnetized}
T_{\rm UM}(z)=\Tb.
\end{equation}
Figure~\ref{fig1} summarizes the main effects of magnetic-gravity filtration on the density and temperature structure of the plasma.

In the left panel, the magnetized density profile computed from Eq.~\eqref{densitypanne} is compared with the corresponding unmagnetized barometric solution. Although both profiles decrease with height because of gravity, the magnetized density exhibits a substantially stronger depletion. This additional reduction is a direct consequence of the progressive formation of the loss cone, which removes from the local distribution function the particles that are mirrored before reaching a given altitude. As a result, the accessible phase-space volume decreases with height, producing densities significantly smaller than those predicted by the purely barometric stratification.

The right panel displays the corresponding temperature profiles. Close to the lower boundary, all temperatures coincide with the boundary Maxwellian value, reflecting the isotropic nature of the distribution function. As the magnetic field expands, magnetic-gravity filtration selectively removes particles with large pitch angles, generating a pronounced temperature anisotropy.

The perpendicular temperature decreases monotonically with height and eventually approaches zero following the asymptotic power-law behaviour derived in the previous section. In contrast, the parallel temperature initially increases above the boundary value, reaches a maximum at an intermediate altitude, and then gradually returns toward the original temperature. This non-monotonic behaviour reflects the competition between the anisotropization induced by magnetic momemtum conservation and the progressive depletion of the accessible velocity-space region at large heights. The total temperature remains bounded between the parallel and perpendicular temperatures, decreases monotonically with altitude, and asymptotically approaches one third of the boundary temperature.

Overall, the figure illustrates how magnetic-gravity filtration simultaneously produces a substantial density depletion and a strong thermal anisotropy within the expanding flux tube. The ordering given by Eq.~\eqref{orderingtemperatures} is clearly visible throughout most of the domain and represents one of the most characteristic signatures of the loss-cone filtering mechanism.

In Figure~\ref{fig1} we presented the solution for a fixed value of the expansion length \(d\). Figure~\ref{fig2} illustrates how the solution
depends on this parameter. Increasing \(d\) shifts the magnetic-field expansion to progressively larger heights, so that the effects of magnetic field become important only at increasingly higher altitudes. This behaviour is clearly visible when comparing the three
rows of Figure~\ref{fig2}. From the top to the bottom row, the differences between the magnetized and unmagnetized density and temperature profiles become progressively smaller over the height range
shown. In the limit of large \(d\), the magnetic field varies only weakly throughout the computational domain, and the magnetized solution gradually approaches the corresponding unmagnetized one, while preserving all the general properties derived analytically in
Section~\ref{sec:General_properties}.

\subsubsection{Location and amplitude of the maximum parallel temperature}
\label{subsubsec:tparallelmax_asymptotic}

Figure~\ref{fig2} shows that the location of the maximum parallel temperature moves to progressively larger heights as the parameter \(d\) increases, i.e. as the magnetic-field expansion becomes less rapid. This behaviour can be understood analytically through the following asymptotic analysis, in which we consider the family of magnetic-field profiles
\begin{equation}
B(z)
=
B_b
\frac{d^{p}}
{\left(z^2+d^2\right)^{p/2}},
\qquad
p\ge2,
\label{eq:generalized_potential_field}
\end{equation}
which reduces to the magnetic field given by Eq.~(\ref{Bpotential}) for \(p=2\).

We consider the region around the maximum parallel temperature. For the typical coronal temperature adopted in the previous section, \(T_b=10^6\,\mathrm{K}\), and for the magnetic-field configuration with \(p=2\), i.e. that given by Eq.~(\ref{Bpotential}), Figs.~\ref{fig1} and \ref{fig2} show that the maximum is located in a region where
\begin{equation}
d\ll z_{\rm max}\ll H.
\label{eq:z_max_range}
\end{equation}
We retain these conditions for \(p\geq2\) below, as an assumption to be checked against the resulting expressions. The left-hand inequality implies \(B(z)/B_b\ll1\). Expanding Eq.~(\ref{paralleltemperaturepanne}) to first order in \(B/B_b\) and \(z/H\) gives
\begin{equation}
\frac{T_\parallel(z)}{T_b}
\simeq
3
-
4\frac{z}{H}
-
\frac{3}{2}\frac{B(z)}{B_b}.
\label{eq:tparallelmaxexp}
\end{equation}
The two correction terms have opposite effects on the local temperature gradient. Since the magnetic field decreases with height, the magnetic contribution becomes progressively less negative and therefore drives an increase in \(T_\parallel\). Gravity, on the other hand, produces a monotonic decrease proportional to \(z/H\).

Differentiating Eq.~(\ref{eq:tparallelmaxexp}) and imposing \(dT_\parallel/dz=0\) yields the condition determining the position of the maximum,
\begin{equation}
-
\left.
\frac{dB}{dz}
\right|_{z=z_{\rm max}}
\simeq
\frac{8\,B_b}{3\,H}.
\label{eq:maxcondition}
\end{equation}

For the magnetic field given by Eq.~(\ref{eq:generalized_potential_field}),
\begin{equation}
-\frac{1}{B_b}\frac{dB}{dz}
=
\frac{
p\,d^{p}\,z
}{
\left(z^2+d^2\right)^{p/2+1}
},
\end{equation}
and substituting this expression into Eq.~(\ref{eq:maxcondition}) gives
\begin{equation}
\frac{
p\,d^{p}\,z_{\rm max}
}{
\left(z_{\rm max}^2+d^2\right)^{p/2+1}
}
\simeq
\frac{8}{3H}.
\label{eq:maxcondition_generalized_potential}
\end{equation}
Using \(d\ll z_{\rm max}\), as given by Eq.~(\ref{eq:z_max_range}), Eq.~(\ref{eq:maxcondition_generalized_potential}) reduces to
\begin{equation}
z_{\rm max}
\simeq
\left(\frac{3p}{8}\right)^{1/(p+1)}
d^{p/(p+1)}
H^{1/(p+1)}.
\label{eq:zmax_generalized_potential}
\end{equation}
For \(p\geq2\), \(\left(3p/8\right)^{1/(p+1)}\approx1\), so \(z_{\rm max}\) indeed approximately satisfies the anticipated conditions in Eq.~(\ref{eq:z_max_range}). For \(p=2\),
\(z_{\rm max}\approx(d^2H)^{1/3}\), so \(z_{\rm max}\) is closer to \(d\) than to \(H\), and \(z_{\rm max}\) increases with the magnetic scale \(d\), in agreement with the numerical results shown in Fig.~\ref{fig2}. Equivalently, for a fixed magnetic-field profile, the
maximum moves closer to the coronal base as \(H\) decreases. Finally, as \(p\) increases, corresponding to a faster decrease of \(B\) with height, \(z_{\rm max}\) approaches \(d\), where \(B(z)\) transitions
from a nearly uniform value at lower heights to a sharply decreasing field.

Finally, substituting Eq.~(\ref{eq:zmax_generalized_potential}) into Eq.~(\ref{eq:tparallelmaxexp}) gives
\begin{equation}
\frac{T_{\parallel,\max}}{T_b}
\simeq
3
-
4
\frac{1+p}{p}
\left(\frac{3p}{8}\right)^{1/(p+1)}
\left(\frac{d}{H}\right)^{p/(p+1)}.
\label{eq:tparallelmax_generalized}
\end{equation}
Therefore, for magnetic fields of the form given by
Eq.~(\ref{eq:generalized_potential_field}), the maximum parallel temperature approaches the gravity-free limit,
\(T_{\parallel,\max}=3T_b\), described by
Eq.~(\ref{asymptoticTgzero}), according to
\begin{equation}
3T_b-T_{\parallel,\max}
\propto
\left(\frac{d}{H}\right)^{p/(p+1)}.
\end{equation}
Increasing \(p\) produces a more rapidly decreasing magnetic field with height, thereby modifying both the location \(z_{\rm max}\) and the value of the maximum temperature, i.e. the rate at which it approaches
the gravity-free limit.

We now validate the analytical estimates against a more accurate numerical calculation for the magnetic-field topology given by Eq.~\eqref{Bpotential}. To this end, we start from the parallel temperature given by Eq.~\eqref{paralleltemperaturepanne}, use the
magnetic field defined by Eq.~\eqref{Bpotential}, differentiate the resulting expression with respect to \(z\), and impose \(dT_\parallel/dz=0\). This yields the following equation for the location of the maximum:
\begin{equation}\label{numericalzmax}
\begin{split}
\frac{\lambda(1+x_{\rm max}^2)}{x_{\rm max}}
+
1
-
2x_{\rm max}^2
&\\
+
\frac{
2x_{\rm max}^3
}{
\sqrt{1+x_{\rm max}^2}
}
\exp\left(
-\frac{\lambda}{x_{\rm max}}
\right)
&=
0.
\end{split}
\end{equation}
where we have introduced the dimensionless variables
\begin{equation}
x\equiv\frac{z}{d},
\qquad
\lambda\equiv\frac{d}{H}.
\end{equation}
Eq.~\eqref{numericalzmax} is a transcendental equation and is solved numerically using the Newton-Raphson method. Once the location of the maximum has been determined, its amplitude can be calculated from Eq.~\eqref{paralleltemperaturepanne} combined with
Eq.~\eqref{Bpotential}. In dimensionless form, this expression reads
\begin{equation}\label{numericalTmax}
\frac{T_{\parallel,\max}}{T_b}
=
\frac{
1-
\dfrac{x_{\rm max}^3}
{(1+x_{\rm max}^2)^{3/2}}
\exp(-\lambda/x_{\rm max})
}{
1-
\dfrac{x_{\rm max}}
{\sqrt{1+x_{\rm max}^2}}
\exp(-\lambda/x_{\rm max})
}.
\end{equation}

In Fig.~\ref{fig3}, we compare the numerical results with the analytical predictions. A good agreement is found for both the location and the amplitude of the maximum, and the agreement improves as \(d\) decreases.
This is because, for smaller values of \(d\), the condition \(z_{\rm max}\ll H\), assumed in the analytical derivation, is increasingly well satisfied. More importantly, the analytical predictions correctly reproduce the dependence on \(d\): the location
of the maximum increases with \(d\), whereas $T_{\parallel,\max}$ decreases.

\section{Multi-temperature case}\label{sec:multi-temperature}

\subsection{Stochastic heating}
\label{sec:Stochastic_heating}

So far, we have considered the case in which the temperature of the Maxwellian distribution at the lower boundary is fixed. The formalism developed in the previous sections is, however, completely general and does not rely on a specific choice of the boundary temperature or density. For the quantitative calculations presented so far, we have adopted a single Maxwellian with a temperature of approximately \(10^6\,\mathrm{K}\), corresponding to a minimal representation of the low corona.

As in our previous studies \cite{Barbieri2023temperature,Barbieri2025c}, the ultimate goal is to model the transition region and the overlying corona. To this end, we now relax the assumption of a fixed boundary temperature and introduce a more realistic boundary condition motivated by observations of the lower transition region.

A large body of observations indicates that the lower transition region is highly dynamic and continuously affected by localized, short-lived heating events (see the observational references discussed in the Introduction). As a consequence, the effective boundary temperature is expected to fluctuate in time rather than remain strictly constant.

Motivated by these observational results, a stochastic heating model was introduced in \cite{Barbieri2023temperature} and subsequently employed in \cite{Barbieri2024b,Barbieri2025,Barbieri2025b,barbieri2026,barbieri2026c}. The aim of this model is to reproduce the temperature fluctuations associated with the heating events.

Let \(T_0\) denote the temperature of the background chromosphere top (approximately \(10^4\,\mathrm{K}\) in the solar case). Over a characteristic time interval \(\tau\), a localized heating event raises the boundary temperature to
\begin{equation}
    T=T_0+\Delta T,
\end{equation}
where the temperature increment \(\Delta T\) is sampled from a probability distribution \(\gamma(T)\).

After the heating phase, the temperature relaxes back to the background value \(T_0\) and remains there for a waiting time \(t_w\). A new heating event then occurs, and the process repeats continuously.

In established coronal-heating models
\citep{Klimchuk_2006,Klimchuk2015}, the heating process is commonly
represented through a prescribed forcing or heating function entering the
fluid equations. The properties of this function determine the energy input
into the plasma and consequently affect macroscopic quantities such as the
density, pressure, and temperature profiles. In the present kinetic model,
stochastic heating is instead introduced through the boundary condition of
the Vlasov equation. Since the description is kinetic, a time-dependent
boundary temperature corresponds to a time-dependent velocity
distribution at the lower boundary, as described above.
Through their subsequent kinetic evolution, the statistical properties of the particle populations injected at the boundary contribute to
shaping the resulting density and temperature profiles.

\subsection{Temporal coarse-graining}
\label{sec:Temporal_coarse-graining}

Previous analytical and numerical studies have shown that the electron relaxation time \(t_{R,e}\) is given by the characteristic crossing time of the system \cite{Barbieri2024b,Barbieri2025b}. More quantitatively,
the relaxation time can be estimated as\cite{Barbieri2024b,Barbieri2025b}
\begin{equation}\label{relaxationtimes}
\begin{aligned}
    t_R &\equiv \min(t_T,t_g), \\
    t_T &= \frac{2L}{v_{T,e}}, \\
    t_g &= \sqrt{\frac{2Lm_e}{gm}}.
\end{aligned}
\end{equation}
where \(2L\) is the loop length, \(v_{T,e}\) is the characteristic electron thermal velocity. Here, \(t_T\) and \(t_{g}\) represent, respectively, the characteristic thermal and total external force crossing times of the system. When the characteristic heating and waiting times, \(\tau\) and \(t_w\), are much shorter than \(t_{R,e}\), the energy injected at the lower boundary cannot be redistributed throughout the plasma during either the heating or the cooling phases \cite{Barbieri2024b,Barbieri2025b}. As a consequence, the plasma does not relax toward thermal equilibrium after each heating event. Instead, the system evolves toward a non-thermal stationary state resulting from the coexistence and mixing of particle populations characterized by different temperatures. The regime considered in the present work is therefore
\begin{equation}
\label{shorttimescalesregime}
\tau,\;t_w\ll t_{R,e}.
\end{equation}

It is important to emphasize that the stationary state considered here does not imply that individual particles remain at a fixed position and experience successive heating events. Particles continuously propagate through the system, with a characteristic crossing time of order \(t_{R,e}\). However, since \(\tau,t_w\ll t_{R,e}\), many heating events occur over a characteristic crossing time. The stationary state refers instead to the statistical distribution resulting from the continuous injection, propagation, and mixing of particle populations associated with different heating events.

Introducing a coarse-graining time scale \(\tilde{t}\), the corresponding coarse-grained distribution function is defined as
\begin{equation}\label{eq:coarse_grained_VDF_def}
\tilde{f}_{\alpha}
=
\frac{1}{\tilde{t}}
\int_{\tilde{t}}
f_{\alpha}\,dt.
\end{equation}

As shown by \cite{Barbieri2025b}, the coarse-grained distribution satisfies a kinetic equation containing the usual Vlasov operator plus an additional contribution arising from the temporal coarse-graining
procedure. In the regime defined by Eq.~\eqref{shorttimescalesregime}, the fluctuations around the coarse-grained distribution scale as
\(\delta f_\alpha = O(\tau/t_{R,\alpha})\), implying that the coarse-graining correction to the Vlasov equation is of second order, \(O[(\tau/t_{R,\alpha})^2]\), and can therefore be neglected. Consequently, the coarse-grained dynamics is governed by the Vlasov equation,
\begin{equation}\label{eq:coarse_grained_evolution}
\frac{\partial \tilde{f}_{\alpha}}
{\partial \tilde{t}}
+
\mathbf{v}\cdot\nabla \tilde{f}_{\alpha}
+
\frac{\mathbf{F}_{\alpha}}{m_{\alpha}}
\cdot
\nabla_{\mathbf{v}}\tilde{f}_{\alpha}
=
0.
\end{equation}

The validity of this approximation, together with the transition to regimes where the coarse-graining correction becomes significant, was investigated in detail in \cite{Barbieri2025b} by means of extensive
particle simulations over a broad range of heating and waiting time scales. Those simulations confirm that, under the condition \eqref{shorttimescalesregime}, the stationary density and temperature profiles are accurately described by the coarse-grained Vlasov equation used in the present work.

Furthermore, \cite{Barbieri2024b,Barbieri2025b} showed that, once the system has reached a stationary state, the fluctuating thermal boundary condition can be temporally coarse-grained over a time interval \(\tilde{t}\) satisfying
\begin{equation}
    \tilde{t} \gg \tau,t_w.
\end{equation}
As discussed above, in the regime defined by Eq.~\eqref{shorttimescalesregime}, the system reaches a stationary state after an initial transient as a consequence of the mixing of the particle populations generated by successive heating events. This behavior has been demonstrated through numerical simulations in \cite{Barbieri2023temperature,Barbieri2024b,Barbieri2025b}.
The coarse-graining interval therefore encompasses many heating events. Under the ergodic assumption, the temporal average over the heated phases can then be replaced by an ensemble average over the temperature distribution \(\gamma(T)\),
\begin{equation}
\begin{aligned}
    &\frac{1}{\tilde{t}_{\rm h}}
    \int_{\tilde{t}_{\rm h}}
    \frac{1}{T(t)^{3/2}}
    \exp\!\left[
    -\frac{m_{\alpha}v^2}{2k_B T(t)}
    \right]dt
    \\
    &\qquad =
    \int_{T_0}^{\infty}
    \frac{\gamma(T)}{T^{3/2}}
    \exp\!\left[
    -\frac{m_{\alpha}v^2}{2k_B T}
    \right]dT .
\end{aligned}
\end{equation}
where \(\tilde{t}_{\rm h}\) is the total time spent in the heated state within the coarse-graining interval. Consequently, once \(\tilde{t}\) is sufficiently long to sample the heating statistics, its explicit time dependence disappears and the coarse-grained distribution at the boundary becomes
\begin{equation}
\label{multitempboundary}
\begin{aligned}
\tilde{f}_{\alpha}(0,v)
={}&
\mathcal{N}_{\alpha}
\Bigg[
A_t
\int_{T_0}^{\infty}
\frac{\gamma(T)}{T^{3/2}}
\exp\!\left(
-\frac{m_{\alpha}v^2}{2k_B T}
\right)
\,dT
\\
&\qquad
+
\frac{1-A_t}{T_0^{3/2}}
\exp\!\left(
-\frac{m_{\alpha}v^2}{2k_B T_0}
\right)
\Bigg],
\end{aligned}
\end{equation}
where \(A_t\) denotes the fraction of time during which a given loop experiences a heating event,
\begin{equation}
    A_t=\frac{\tau}{\tau+t_w},
\end{equation}
and the normalization constant is
\begin{equation}
    \mathcal{N}_{\alpha}
    =
    n_0
    \left(
    \frac{m_{\alpha}}
         {2\pi k_B}
    \right)^{3/2}.
\end{equation}

Applying the Liouville mapping to Eq.~\eqref{multitempboundary}, the distribution function throughout phase space becomes
\begin{equation}
\label{VDFphasespace}
\begin{aligned}
\tilde{f}_{\alpha}(z,v)
=
\chi(z,\mathbf{v})\,\mathcal{N}_{\alpha}
\Bigg[
&A_t
\int_{T_0}^{\infty}
\frac{\gamma(T)}{T^{3/2}}
\exp\!\left(
-\frac{\mathcal{H}_{\alpha}}{k_B T}
\right)
\,dT
\\
&+
\frac{1-A_t}{T_0^{3/2}}
\exp\!\left(
-\frac{\mathcal{H}_{\alpha}}{k_B T_0}
\right)
\Bigg].
\end{aligned}
\end{equation}

As discussed below Eq.~\eqref{eq:coarse_grained_VDF_def}, in the regime 
defined by Eq.~\eqref{shorttimescalesregime}, the fluctuations \(\delta f_\alpha\) around the coarse-grained distribution are negligible. After the initial transient, the coarse-grained distribution reaches the statistically stationary state discussed above, so that we can identify \(\tilde{f}_\alpha\simeq f_\alpha\) within the accuracy of the present approximation.

We stress that, by representing each individual heating event with a Maxwellian distribution, the present model deliberately adopts a conservative assumption regarding the non-thermal content of the heated particle population. Importantly, this assumption applies only to the distribution associated with each heating event at the lower boundary and does not imply that the plasma remains in thermal equilibrium throughout the atmosphere. Indeed, gravitational filtering and magnetic-moment conservation modify the velocity distribution as particles propagate upward, naturally producing non-Maxwellian features and temperature anisotropies.

Moreover, under stochastic heating, the coarse-grained distribution is itself a superposition of Maxwellian populations characterized by different temperatures and is therefore not, in general, a single Maxwellian equilibrium distribution. The presence of suprathermal particles in the distributions associated with individual heating events are enhanced by the gravitational filtering which  therefore strengthen the kinetic effects investigated here.

As in the single-temperature case, only the region of phase space compatible with the conservation of total energy, Eq.~\eqref{hamiltonian}, and magnetic moment, Eq.~\eqref{magneticmoment}, can be populated. This is accounted for by the characteristic function \(\chi(z,\mathbf{v})\), defined in Eq.~\eqref{velocityfilter}. Although \(\chi(z,\mathbf{v})\) appears outside the temperature integral in Eq.~\eqref{VDFphasespace}, its effect is fully retained when calculating the moments of the distribution function. Indeed, as shown in the following sections, the moments are obtained by integrating over velocity space, with \(\chi(z,\mathbf{v})\) restricting the integration to the dynamically accessible region. Owing to the linearity of the integration, the velocity-space integration can be performed separately for each temperature component. The resulting moment of each Maxwellian component depends on its temperature \(T\) and therefore remains inside the subsequent integral over \(T\). The total moment is then obtained by superposing these single-temperature contributions according to the statistical weight \(\gamma(T)\). 

\subsection{Multi-temperature moments}
\label{sec:Multi-temperature_moments}

Using the multi-temperature distribution function,
Eq.~\eqref{VDFphasespace}, and computing the corresponding moments by exploiting the results obtained for a single Maxwellian population, the density becomes
\begin{equation}
\label{density_multiT}
\begin{aligned}
n_{\alpha}(z)
={}&
n_0
\Bigg[
A_t
\int_{T_0}^{+\infty}
dT\,\gamma(T)
e^{-V_{\alpha}(z)/(k_B T)}
\\
&\qquad\qquad\times
\left(
1-\mathcal{C}(z)e^{-\as_{\alpha}(z,T)}
\right)
\\
&\qquad
+
(1-A_t)
e^{-V_{\alpha}(z)/(k_B T_0)}
\\
&\qquad\qquad\times
\left(
1-\mathcal{C}(z)e^{-\as_{\alpha}(z,T_0)}
\right)
\Bigg].
\end{aligned}
\end{equation}
As in the single-temperature case, we show that \(\phis=0\) remains the unique solution satisfying local charge neutrality (see Appendix \ref{appendix:multitemp_uniqueness}). Since \(\phis=0\), the total potential is the same for both species and is still given by Eq.~\eqref{totalpotential}. Therefore, the density profile becomes
\begin{equation}
\label{densitymultipanne}
\begin{aligned}
n_M(z)
={}&
n_0
\Bigg[
A_t
\int_{T_0}^{+\infty}
dT\,\gamma(T)\,
e^{-z/\HT}
\\
&\qquad\qquad\times
\left(
1-\mathcal{C}(z)e^{-\aT}
\right)
\\
&\qquad
+
(1-A_t)\,
e^{-z/\HTo}
\\
&\qquad\qquad\times
\left(
1-\mathcal{C}(z)e^{-\aTo}
\right)
\Bigg].
\end{aligned}
\end{equation}
where $H_T$ and $H_{T_0}$ are the gravitational scale heights for temperatures $T$ and $T_0$, respectively. We also simplify the notation by writing $\as(z, T)$ and $\as(z, T_0)$ simply $\aT$ and $\aTo$, respectively.
The parallel temperature is given by
\begin{equation}
\label{paralleltemperaturepanne2}
\begin{aligned}
T_{\parallel,M}(z)
={}&
\frac{n_0}{n(z)}
\Bigg[
A_t
\int_{T_0}^{+\infty}
dT\,\gamma(T)\,T\,
e^{-z/\HT}
\\
&\qquad\qquad\times
\left(
1-\mathcal{C}^3(z)e^{-\aT}
\right)
\\
&\qquad
+
(1-A_t)\,
T_0\,
e^{-z/\HTo}
\\
&\qquad\qquad\times
\left(
1-\mathcal{C}^3(z)e^{-\aTo}
\right)
\Bigg].
\end{aligned}
\end{equation}
and the perpendicular temperature is
\begin{equation}
\label{perpendiculartemperaturepanne}
\begin{aligned}
T_{\perp,M}(z)
={}&
\frac{n_0}{n(z)}
\Bigg[
A_t
\int_{T_0}^{+\infty}
dT\,\gamma(T)\,T\,
e^{-z/\HT}
\\
&\qquad\qquad\times
\Biggl(
1
-
\mathcal{C}(z)
\left[
\frac{3}{2}
+
\aT
\right]
e^{-\aT}
\\
&\qquad\qquad\qquad
+
\frac{1}{2}
\mathcal{C}^3(z)e^{-\aT}
\Biggr)
\\
&\qquad
+
(1-A_t)\,
T_0\,
e^{-z/\HTo}
\\
&\qquad\qquad\times
\Biggl(
1
-
\mathcal{C}(z)
\left[
\frac{3}{2}
+
\aTo
\right]
e^{-\aTo}
\\
&\qquad\qquad\qquad
+
\frac{1}{2}
\mathcal{C}^3(z)e^{-\aTo}
\Biggr)
\Bigg].
\end{aligned}
\end{equation}
while the total temperature is obtained via Eq. \eqref{totaltemperaturedef}.

The above expressions show that the density and temperature profiles are obtained as weighted superpositions of Maxwellian populations generated at different temperatures. 

In the limit of a uniform magnetic field, $ \mathcal{C}(z)\rightarrow 0$, the effects of magnetic-gravity filtration vanish and the density and temperature profiles reduce to those obtained for the unmagnetized case \citep{Barbieri2024b}. The density becomes
\begin{equation}
\label{densityunmagnetizedmulti}
\begin{aligned}
n_{UM,M}(z)
={}&
\nb
\Bigg[
A_t
\int_{T_0}^{\infty}
\gamma(T)\,
e^{-z/\HT}
\,dT
\\
&\qquad
+
(1-A_t)\,
e^{-z/\HTo}
\Bigg].
\end{aligned}
\end{equation}
while the kinetic temperature is
\begin{equation}
\label{temperatureunmagnetizedmulti}
\begin{aligned}
T_{UM,M}(z)
={}&
\Bigg[
A_t
\int_{T_0}^{+\infty}
T\,\gamma(T)
e^{-z/\HT}
\,dT
\\
&\qquad
+
(1-A_t)\,
T_0
e^{-z/\HTo}
\Bigg]
\\
&\times
\Bigg[
A_t
\int_{T_0}^{+\infty}
\gamma(T)
e^{-z/\HT}
\,dT
\\
&\qquad
+
(1-A_t)\,
e^{-z/\HTo}
\Bigg]^{-1}.
\end{aligned}
\end{equation}

As shown by \citet{Barbieri2023temperature,Barbieri2024b,Barbieri2025b}, while the density decreases with height, the temperature increases. Physically, this occurs because particles originating from the high-temperature tail of the boundary distribution can penetrate further into the gravitational potential, progressively increasing the mean energy of the particle population at large heights.

The effects of magnetic-field expansion on these profiles are discussed in the following.

\subsection{General properties of the multi-temperature solution}
\label{sec:General_properties_multi-temperature}

Here \(\aT\) is still given by Eq.~\eqref{apanne}. As in the single-temperature case, Eq.~\eqref{propertiesc} together with the condition \(\aT>0\) implies that Eq.~\eqref{propertiesac} remains valid. Therefore,
\begin{equation}
n_M(z)
<
n_{UM}(z),
\end{equation}
for every \(z>0\). Thus, magnetic-gravity filtration always reduces the density relative to the corresponding unmagnetized multi-temperature atmosphere.

For each temperature component, the difference between the kernels entering the parallel and perpendicular temperatures is positive. Indeed,
\begin{equation}
\begin{aligned}
&
1-\mathcal{C}^3(z)e^{-\aT}
\\
&\quad-
\left[
1
-
\mathcal{C}(z)
\left(
\frac32+\aT
\right)
e^{-\aT}
+
\frac12\mathcal{C}^3(z)e^{-\aT}
\right]
\\[3pt]
={}&
\mathcal{C}(z)e^{-\aT}
\left[
\frac32+\aT
-\frac32\mathcal{C}^2(z)
\right].
\end{aligned}
\end{equation}
Since Eq.~\eqref{propertiesc} holds and \(\aT>0\), the quantity in square brackets is strictly positive. Therefore,
\begin{equation}
T_{\parallel,M}(z)
>
T_{\perp,M}(z).
\end{equation}

The total temperature is defined by Eq.~\eqref{totaltemperaturedef}. It then follows immediately that
\begin{equation}
T_M(z)-T_{\perp,M}(z)
=
\frac{
T_{\parallel,M}(z)-T_{\perp,M}(z)
}{3}
>0,
\end{equation}
and
\begin{equation}
T_{\parallel,M}(z)-T_M(z)
=
\frac{2}{3}
\left[
T_{\parallel,M}(z)-T_{\perp,M}(z)
\right]
>0.
\end{equation}

Hence the multi-temperature solution satisfies the robust ordering
\begin{equation}\label{hierarckyT}
T_{\perp,M}(z)
<
T_M(z)
<
T_{\parallel,M}(z).
\end{equation}
This ordering is independent of the detailed shape of the temperature distribution $\gamma(T)$ and follows solely from the loss-cone structure produced by magnetic moment conservation.

\subsection{The regime of rare and intense heating events}
\label{sec:intense_heating_events}

In this section, we analyze the relative importance of the different temperature components in the physically relevant regime
\begin{equation}\label{physicalregime}
\begin{aligned}
A_t &\ll 1, \\
\Delta T &= \int_{T_0}^{\infty} (T-T_0)\gamma(T)\,dT
\sim 10^{6}\,\mathrm{K} \gg T_0.
\end{aligned}
\end{equation}

Physically, this limit corresponds to heating events that are rare in time, with a temperature distribution extending to values much larger than the background chromospheric temperature. This is the regime in which the unmagnetized multi-temperature model investigated by
\citet{Barbieri2023temperature,Barbieri2024b} reproduces the observed solar density and temperature profiles, including a million-degree corona. If the mean temperature increase \(\Delta T\) is instead much smaller than \(10^6\,\mathrm{K}\), the model does not produce sufficiently energetic particles to reproduce the coronal temperature.

Importantly, the condition \(\Delta T\sim10^6\,\mathrm{K}\) does not imply that individual heating events must produce temperature increases of this magnitude. Rather, \(\Delta T\) represents the mean temperature increment of the complete distribution \(\gamma(T)\). For a monotonically decreasing distribution starting from \(T_0\), low-temperature heating events are the most probable, whereas events reaching temperatures of order
\(10^6\,\mathrm{K}\) or higher belong to the tail of the distribution and are correspondingly less frequent. Nevertheless, these rare
high-temperature populations contain particles with sufficient energy to climb higher in the gravitational potential and therefore become progressively more important with increasing altitude through gravitational filtering. Consequently, the million-degree corona is produced by the preferential selection of the high-temperature tail of \(\gamma(T)\), rather than by requiring all heating events to reach coronal temperatures.

The specific form of \(\gamma(T)\) and its observational motivation are discussed in
Sect.~IV.G. There, we adopt an exponentially decreasing distribution, motivated by recent observational and numerical studies, for which lower-temperature heating events are more frequent, while progressively hotter events become increasingly rare.

In the unmagnetized case, the density and temperature are given by Eqs.~\eqref{densityunmagnetizedmulti} and \eqref{temperatureunmagnetizedmulti}, respectively.

In the limit \(A_t\ll1\), the hot population contributes only a small fraction of the local density near the lower boundary. However, the ratio between the density contribution of the hot population and that of the background population increases with height. This can be seen by comparing the contribution of a temperature component \(T>T_0\) with that of the background component at \(T_0\).

For the density, this ratio is
\begin{equation}
\begin{aligned}
\mathcal{R}_n(z,T)
={}&
\frac{A_t}{1-A_t}\,
\gamma(T)
\\
&\times
\exp\!\left[
\frac{z}{\HTo}
-
\frac{z}{\HT}
\right]
\\
&\times
\frac{
1-\mathcal{C}(z)e^{-\aT}
}{
1-\mathcal{C}(z)e^{-\aTo}
}.
\end{aligned}
\end{equation}
For every \(T>T_0\), the exponential factor increases with height because the hotter components possess larger gravitational scale heights $\HT$.
Consequently, the colder background population, characterized by the shorter scale height $\HTo$ decreases much more rapidly with altitude than the hotter components. Even if the hot population represents only a very small fraction of the particles at the lower boundary (\(A_t\ll1\)), its much slower exponential decay progressively compensates for its smaller abundance, so that sufficiently hot components eventually dominate the density at large heights.

Moreover, under the general assumptions introduced in
Section~\ref{sec:Spatial_limits}, the loss-cone factor satisfies Eq. \eqref{asymptotica} and Eq. \eqref{asymptoticaC}. Therefore,
\begin{equation}
\frac{
1-\mathcal{C}(z)e^{-\aT}
}{
1-\mathcal{C}(z)e^{-\aTo}
}
\simeq
\frac{\HTo}{\HT}
=
\frac{T_0}{T}.
\end{equation}
Thus, at large heights the magnetic field contributes only through a temperature-dependent algebraic prefactor that is independent of height. Consequently, it cannot compete with the exponential enhancement produced by the different gravitational scale heights.
Magnetic moment conservation therefore modifies only the relative amplitude of the contribution of each temperature component, while leaving unchanged the fundamental effect of gravitational filtering, namely
the progressive dominance of the hottest populations at large altitudes. These predictions will be confirmed numerically by the profiles presented in Section~\ref{sec:temperature_density_multi}.

\subsection{Asymptotic behaviour of the hot component}
\label{sec:Asymptotic_behaviour}

We now consider the limiting case \(A_t=1\), in which only the hot multi-temperature component is retained. This limiting case is particularly useful because gravitational filtering causes the hotter temperature components to decrease more slowly with height than the
cooler ones. As a result, their relative contribution to the local particle density and to the velocity moments progressively increases at large heights.

Under the general assumptions introduced in
Section~\ref{sec:Spatial_limits}, the asymptotic relations given by Eqs.~\eqref{asymptotica} and \eqref{asymptoticaC} hold for every temperature component. Substituting these expressions into
Eq.~\eqref{densitymultipanne} with \(A_t=1\) yields
\begin{equation}
\begin{split}
n_{\rm hot}(z)
\simeq\;
& n_0
\frac{z\,B(z)}{\Bb}
\\
& \times
\int_{T_0}^{+\infty}
dT\,\gamma(T)
\frac{1}{\HT}
\exp\!\left(
-\frac{z}{\HT}
\right).
\end{split}
\end{equation}
Thus, the combined effect of energy and magnetic-moment conservation introduces the same algebraic depletion factor proportional to \(zB(z)\) as in the single-temperature case.

For comparison, the corresponding unmagnetized density is
\begin{equation}
n_{{\rm UM},\rm hot}(z)
=
n_0
\int_{T_0}^{+\infty}
dT\,\gamma(T)
\exp\!\left(
-\frac{z}{\HT}
\right).
\end{equation}
Therefore,
\begin{equation}
\frac{n_{\rm hot}(z)}
{n_{{\rm UM},\rm hot}(z)}
\simeq
\frac{z\,B(z)}{\Bb}
\,
\frac{
\int_{T_0}^{+\infty}
dT\,\gamma(T)\,\HT^{-1}e^{-z/\HT}
}{
\int_{T_0}^{+\infty}
dT\,\gamma(T)e^{-z/\HT}
}.
\end{equation}
which shows that there is a larger filtration of the hotter particles through the $\HT^{-1}$ term at the numerator. Consequently,

\begin{equation}
\frac{n_{\rm hot}(z)}
{n_{{\rm UM},\rm hot}(z)}
\propto
zB(z).
\end{equation}

If the magnetic field satisfies the asymptotic power law given by Eq.~\eqref{magneticpowerlaw}, the above ratio scales as
\begin{equation}
\frac{n_{\rm hot}(z)}
{n_{{\rm UM},\rm hot}(z)}
\propto
z^{1-p},
\end{equation}
recovering, for the particular magnetic-field profile given by Eq.~\eqref{Bpotential} (\(p=2\)), the previously derived \(z^{-1}\) behaviour.

Thus, although the absolute density profile remains controlled by gravitational filtering through the temperature integral, magnetic moment conservation introduces an additional universal algebraic depletion
proportional to \(zB(z)\). This result is the direct analogue of the single-temperature asymptotic behaviour and is independent of the specific form of the boundary temperature distribution \(\gamma(T)\). 

The parallel temperature is
\begin{equation}
\label{tparallelA1}
\begin{split}
T_{\parallel,\Mhot}(z)
={}&
\frac{
\displaystyle
\int dT\,\gamma(T)\,T\,
e^{-z/\HT}
\left[
1-\mathcal{C}^3(z)e^{-\aT}
\right]
}{
\displaystyle
\int dT\,\gamma(T)\,
e^{-z/\HT}
\left[
1-\mathcal{C}(z)e^{-\aT}
\right]
}.
\end{split}
\end{equation}
Using the asymptotic expansion
\begin{equation}
1-\mathcal{C}^3(z)e^{-\aT}
\simeq
\aT,
\end{equation}
together with Eqs.~\eqref{asymptotica} and
\eqref{asymptoticaC}, Eq.~\eqref{tparallelA1} becomes
\begin{equation}
\label{TparallelMhot}
T_{\parallel,\Mhot}(z)
\simeq
\frac{
\displaystyle
\int dT\,\gamma(T)\,
e^{-z/\HT}
}{
\displaystyle
\int dT\,\gamma(T)\,T^{-1}
e^{-z/\HT}
}.
\end{equation}

The perpendicular temperature is
\begin{equation}
\label{TperpA1}
T_{\perp,\Mhot}(z)
=
\frac{
\int dT\,\gamma(T)\,T\,
e^{-z/\HT}
\,\mathcal{K}_{\perp}(z,T)
}{
\int dT\,\gamma(T)\,
e^{-z/\HT}
\left[
1-\mathcal{C}(z)e^{-\aT}
\right]
},
\end{equation}
where
\begin{equation}
\label{eqKperp}
\begin{split}
\mathcal{K}_{\perp}(z,T)
={}&
1
-
\mathcal{C}(z)
\left(
\frac32+\aT
\right)
e^{-\aT}
\\
&+
\frac12
\mathcal{C}^3(z)e^{-\aT}.
\end{split}
\end{equation}
The asymptotic expansion of the kernel is
\begin{equation}
\mathcal{K}_{\perp}(z,T)
\simeq
\frac{\aT^2}{2}.
\end{equation}
Substituting this expression together with Eqs.~\eqref{asymptotica} and \eqref{asymptoticaC} into
Eq.~\eqref{TperpA1} yields
\begin{equation}
T_{\perp,\Mhot}(z)
\simeq
\frac{
\displaystyle
\int dT\,\gamma(T)\,T\,
e^{-z/\HT}
\frac12
\left(
\frac{z}{\HT}
\frac{B(z)}{\Bb}
\right)^2
}{
\displaystyle
\int dT\,\gamma(T)\,
e^{-z/\HT}
\left(
\frac{z}{\HT}
\frac{B(z)}{\Bb}
\right)
}.
\end{equation}
The magnetic-field-dependent factors cancel except for one power, yielding
\begin{equation}
T_{\perp,\Mhot}(z)
\simeq
\frac{z\,B(z)}{2\,\Bb}
\,
\frac{
\displaystyle
\int dT\,\gamma(T)\,\HT^{-1}
e^{-z/\HT}
}{
\displaystyle
\int dT\,\gamma(T)\,T^{-1}
e^{-z/\HT}
}.
\end{equation}
Hence,
\begin{equation}
T_{\perp,\Mhot}(z)
\propto
zB(z).
\end{equation}
If the magnetic field satisfies the asymptotic power law given by Eq.~\eqref{magneticpowerlaw}, then
\begin{equation}
T_{\perp,\Mhot}(z)
\propto
z^{1-p},
\end{equation}
recovering the \(z^{-1}\) behaviour for the magnetic-field profile given by Eq.~\eqref{Bpotential}.

Thus, the perpendicular temperature of the hot component exhibits the same asymptotic scaling as in the single-temperature solution. This result depends only on the asymptotic geometry of the loss cone and is
independent of the detailed form of the temperature distribution \(\gamma(T)\).

\begin{figure*}
\centering
\includegraphics[width=0.99\textwidth]{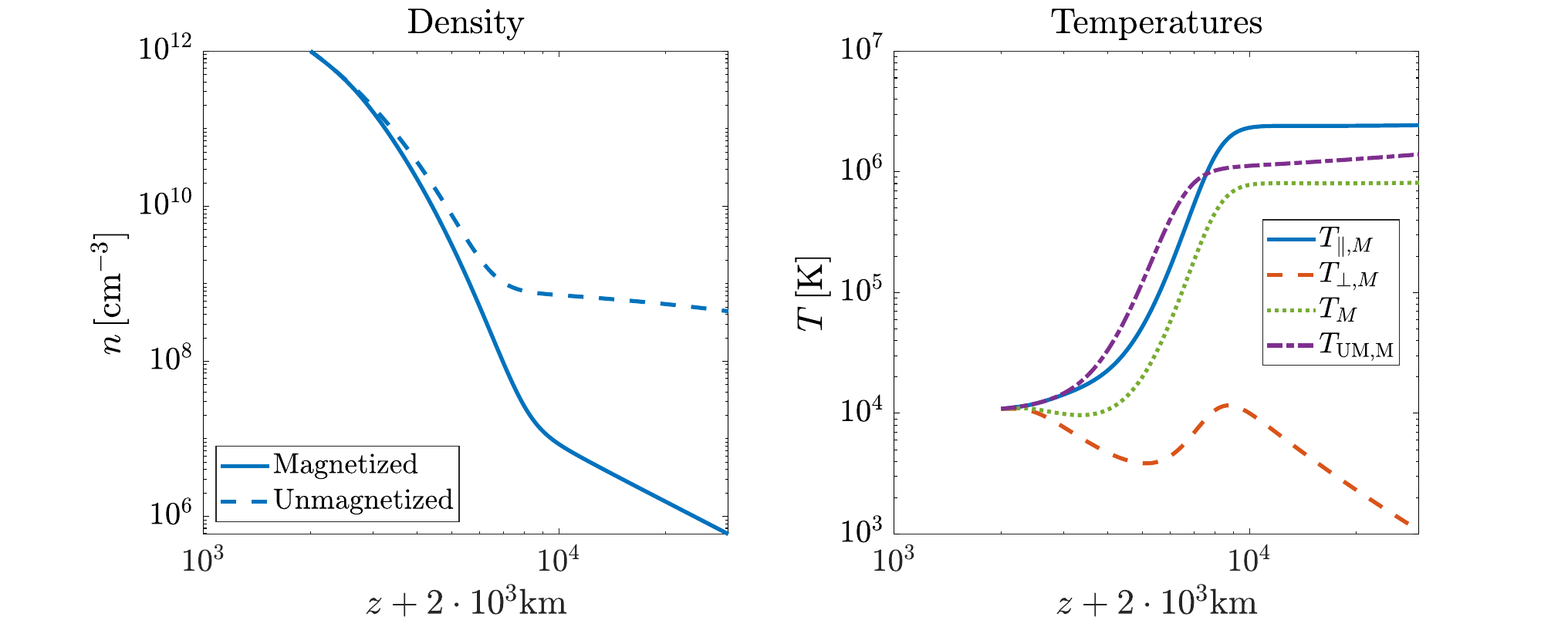}
\caption{
Same as Figure~\ref{fig1}, but for the multi-temperature solution described in Section~\ref{sec:multi-temperature}. The density profile \(n_M(z)\) is computed from Eq.~\eqref{densitymultipanne} and compared with the corresponding unmagnetized profile given by Eq.~\eqref{densityunmagnetizedmulti}. The parallel, perpendicular, and total temperature profiles are computed from Eqs.~\eqref{paralleltemperaturepanne2}, \eqref{perpendiculartemperaturepanne}, and \eqref{totaltemperaturedef}, respectively, and are compared with the corresponding unmagnetized temperature profile given by Eq.~\eqref{temperatureunmagnetizedmulti}. The colour scheme, line styles, and legend conventions are identical to those adopted in Figure~\ref{fig1}. The parameters are \(d=10^3\,\mathrm{km}\), \(n_0=10^{12}\,\mathrm{cm^{-3}}\), \(T_0=10^4\,\mathrm{K}\), \(A_t= 10^{-3}\), and \(\Delta T= 10^6\mathrm{K}\). The density normalization is chosen so that the unmagnetized atmosphere has \(n\simeq10^9\,\mathrm{cm^{-3}}\) and \(T\simeq10^6\,\mathrm{K}\) at the coronal base (\(z\simeq10^4\,\mathrm{km}\)), consistently with Figure~\ref{fig1}.
}
\label{fig4}
\end{figure*}

The total temperature is defined as
\begin{equation}
T_{\hot}(z)
=
\frac{
T_{\parallel,\Mhot}(z)
+
2\,T_{\perp,\Mhot}(z)
}{3}.
\end{equation}

Since \(T_{\perp,\Mhot}(z)\) vanishes asymptotically as
\(zB(z)\), whereas \(T_{\parallel,\Mhot}(z)\) remains controlled by the high-temperature components selected through gravitational filtering, the asymptotic ordering is
\begin{equation}
T_{\perp,\Mhot}(z)
\ll
T_{\hot}(z)
<
T_{\parallel,\Mhot}(z).
\end{equation}
Accordingly,
\begin{equation}
T_{\hot}(z)
\simeq
\frac{
T_{\parallel,\Mhot}(z)
}{3},
\qquad
z\rightarrow\infty.
\end{equation}

The physical interpretation is the following. At large heights, gravitational filtering progressively favours the hottest temperature components, since they possess the largest gravitational scale heights. At the same time, the combined conservation of energy and magnetic
moment continuously reduces the fraction of particle trajectories that remain magnetically connected to the lower boundary. Consequently, the accessible region of velocity space becomes increasingly concentrated
around the magnetic-field direction. The parallel and total temperatures are therefore mainly determined by
the surviving hot population, whereas the perpendicular second velocity moment is progressively depleted by the widening loss cone.

For comparison, the corresponding unmagnetized multi-temperature solution is obtained by setting \(A_t=1\) in
Eq.~\eqref{temperatureunmagnetizedmulti}, yielding
\begin{equation}
\label{TMhot}
T_{\Mhot}(z)
=
\frac{
\displaystyle
\int_{T_0}^{+\infty}
dT\,\gamma(T)\,
T\,
e^{-z/\HT}
}{
\displaystyle
\int_{T_0}^{+\infty}
dT\,\gamma(T)\,
e^{-z/\HT}
}.
\end{equation}

Both \(T_{\Mhot}(z)\) and \(T_{\parallel,\Mhot}(z)\), given by Eq.~\eqref{TparallelMhot}, are controlled by the high-temperature tail selected through gravitational filtering. Their exact values differ,
since the temperature enters the corresponding weighted integrals in different ways. Nevertheless, if the large-height integrals are dominated by a sufficiently narrow range of temperatures, the two quantities are expected to be of the same order,
\begin{equation}
T_{\parallel,\Mhot}(z)
\sim
T_{\Mhot}(z).
\end{equation}
Consequently,
\begin{equation}
T_{\perp,\Mhot}(z)
\ll
T_{\Mhot}(z)
\sim
T_{\parallel,\Mhot}(z),
\qquad
z\rightarrow\infty,
\end{equation}
while
\begin{equation}
T_{\hot}(z)
\simeq
\frac{1}{3}
T_{\parallel,\Mhot}(z)
\sim
\frac{1}{3}
T_{\Mhot}(z).
\end{equation}

The expected asymptotic ordering therefore becomes
\begin{equation}
T_{\perp,\Mhot}(z)
\ll
T_{\hot}(z)
<
T_{\parallel,\Mhot}(z)
\sim
T_{\Mhot}(z).
\end{equation}
\begin{figure*}
\centering
\includegraphics[width=0.99\textwidth]{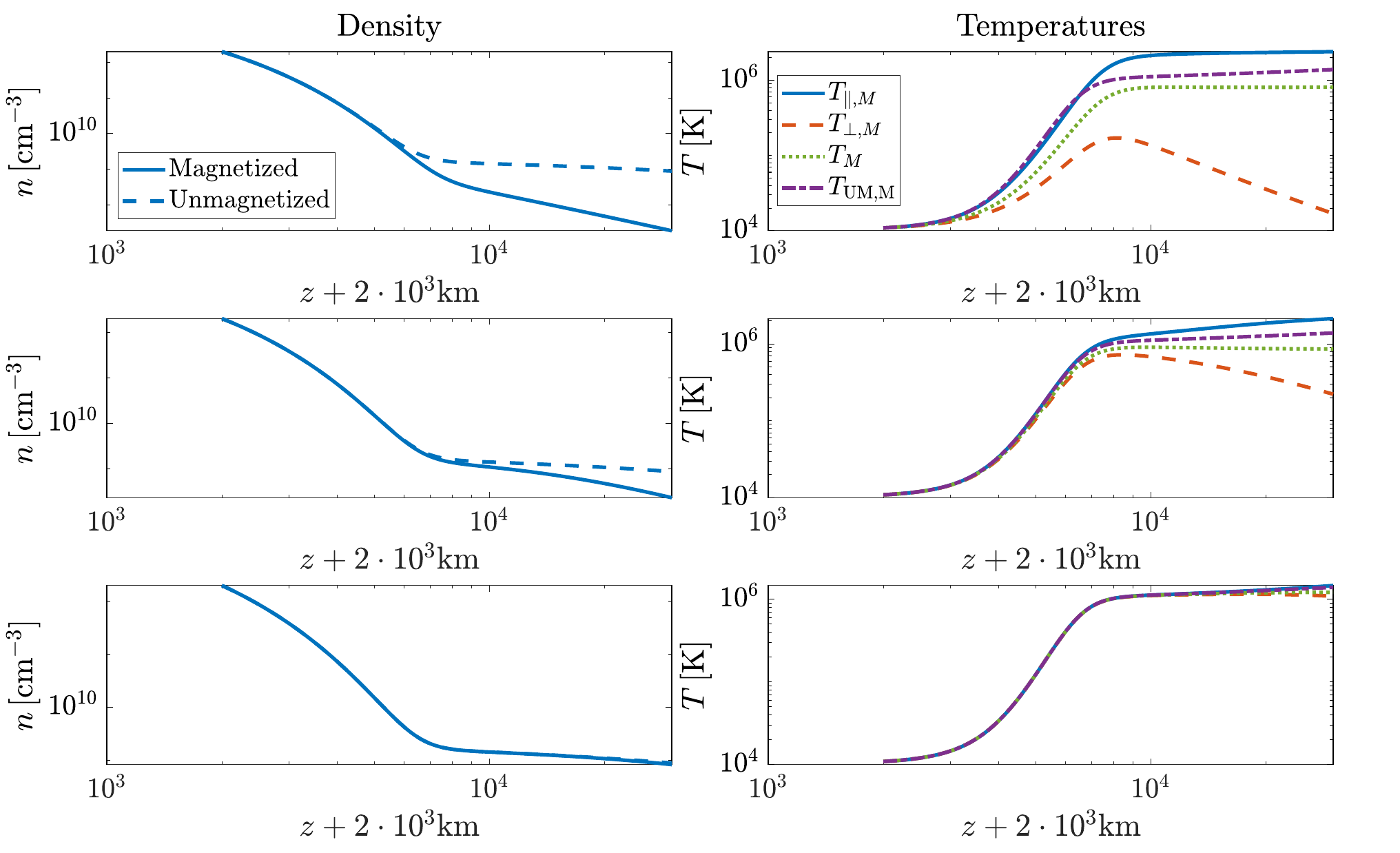}
\caption{
Comparison of the multi-temperature solutions for different magnetic-field expansion lengths \(d\). The rows correspond to the same values of d as in Figure~\ref{fig2}, in the same top-to-bottom order. In each row, the left panel shows the magnetized and unmagnetized density profiles computed from Eqs.~\eqref{densitymultipanne} and \eqref{densityunmagnetizedmulti}, respectively, while the right panel shows the corresponding parallel, perpendicular, total, and unmagnetized temperature profiles computed from Eqs.~\eqref{paralleltemperaturepanne2}, \eqref{perpendiculartemperaturepanne}, \eqref{totaltemperaturedef}, and \eqref{temperatureunmagnetizedmulti}, respectively. The colour scheme, line styles, legend conventions, and all other parameters are identical to those adopted in Figure~\ref{fig2}.
}
\label{fig5}
\end{figure*}

\subsection{Temperature and density profiles} \label{sec:temperature_density_multi}

We now present the density and temperature profiles, highlighting the anisotropy induced by magnetic moment conservation and comparing the magnetized solution with the corresponding unmagnetized case. Throughout this section, we consider the physical regime defined by Eq.~\eqref{physicalregime} and adopt the same magnetic-field model used in Section~\ref{subsubsec:densitytempsingle} for the single-temperature case and given by Eq. \eqref{Bpotential}.

Regarding the distribution of heating events, it was shown by \citet{Barbieri2025c} that, once the above constraints given by Eq. \eqref{physicalregime}, the detailed functional form of $\gamma(T)$ has only a minor influence on the resulting density and temperature profiles. Throughout this work we therefore adopt the exponential
distribution
\begin{equation}
\label{exponentialincrements}
\gamma(T)
=
\frac{1}{\Delta T}
\exp
\left[
-\frac{T-T_0}{\Delta T}
\right],
\qquad
T>T_0.
\end{equation}
As discussed by \citet{barbieri2026}, this choice is motivated by recent observational and numerical studies \citep{Huang2023,Dolliou2023,Dolliou_2024,Dolliou2025}, which indicate that lower-temperature heating events are more frequent, whereas progressively hotter events become increasingly rare. Moreover, as already discussed below Eq.~\eqref{physicalregime} in Sect.~\ref{sec:intense_heating_events}, the condition \(\Delta T\sim10^6\,{\rm K}\) does not imply that all heating events produce a temperature increase of this magnitude. For the exponential distribution adopted here, the probability density is largest for small temperature increments and decreases monotonically with increasing \(T-T_0\). Heating events reaching coronal temperatures therefore populate the progressively less probable high-temperature tail of \(\gamma(T)\).

Figure~\ref{fig4} summarizes the main properties of the
multi-temperature solution discussed above. The left panel compares the density profiles obtained in the magnetized and unmagnetized cases. In the present example, the characteristic expansion length \(d\) is chosen to be smaller than the thickness of the transition region. Consequently, magnetic-field expansion already affects the plasma properties within the lower atmosphere. The magnetized density is therefore systematically smaller than its unmagnetized counterpart, with the difference becoming progressively more pronounced with height. This additional depletion results from the combined conservation of energy and magnetic moment, which progressively restricts the populated region of velocity space through the formation of a loss cone. As predicted by the asymptotic analysis, the discrepancy between the two density profiles increases with height, reflecting the additional depletion introduced by magnetic moment conservation on top of gravitational filtration.

The right panel displays the corresponding temperature profiles. Near the lower boundary, all temperatures remain close to the chromospheric reference value \(T_0\), since the contribution of the hot population is
still negligible. As the height increases, gravitational filtering progressively selects the hotter components of the boundary temperature
distribution because of their larger gravitational scale heights, leading to a substantial increase of the total temperature with respect to the lower atmosphere. In the unmagnetized case, this mechanism produces the monotonic increase of \(T_{UM,M}\), which approaches the
characteristic temperature of the surviving hot population.

The magnetized solution is governed by the same gravitational filtering mechanism. However, because the magnetic field already undergoes an appreciable expansion across the transition region, the combined effect of conservation of energy and magnetic moment modifies the temperature profiles already at low heights. At coronal altitudes, the parallel temperature rises above the unmagnetized solution and becomes the largest temperature component. The perpendicular temperature, on the other hand, exhibits a non-monotonic dependence on height: after reaching a modest maximum, it decreases rapidly as the loss cone progressively removes particles with large perpendicular velocities. The total temperature remains bounded between the parallel and perpendicular temperatures at all heights, consistently with the general ordering derived in Section~\ref{sec:General_properties_multi-temperature}. At sufficiently large heights it approaches approximately one third of the parallel temperature, in agreement with the asymptotic analysis.

Figure~\ref{fig5} illustrates how the solution depends on the expansion parameter \(d\). Compared with the reference case shown in Figure~\ref{fig4}, progressively larger values of \(d\) shift the magnetic-field expansion to increasingly larger heights.

The first row corresponds to an expansion length that is still smaller than the thickness of the transition region, although larger than in the reference case. Consequently, the magnetic moment still modifies both the density and, more noticeably, the temperature profiles throughout the transition region. However, these deviations from the corresponding unmagnetized solution are less pronounced than those shown in
Figure~\ref{fig4}. Nevertheless, all the general properties derived analytically in Section~\ref{sec:General_properties_multi-temperature} remain
satisfied.

The second row corresponds to an expansion length comparable to the thickness of the transition region. In this case, the magnetic field varies only weakly throughout the lower atmosphere, and the effects of
magnetic moment conservation become negligible within the transition region, remaining significant only in the low corona, similarly to the single-temperature case shown in Figure~\ref{fig1}.

Finally, the third row corresponds to an expansion length much larger than the transition region size. In this limit, the magnetic field remains nearly constant over the heights considered, and the
magnetized and unmagnetized solutions become almost indistinguishable from the chromosphere up to the low corona. Only at the largest heights shown do small deviations begin to appear, as the cumulative effect of
magnetic-field expansion gradually becomes appreciable.

These results demonstrate that the parameter \(d\) primarily controls the height at which the combined effect of conservation of energy and magnetic moment becomes important. A smaller value of \(d\) concentrates the magnetic-field expansion within the lower atmosphere, causing the loss-cone anisotropy to develop already inside the transition region. Conversely, increasing \(d\) shifts the onset of magnetic effects to progressively larger heights while leaving the general properties of the solution unchanged. In particular, the temperature hierarchy given by Eq.~\eqref{hierarckyT} remains satisfied for all values of \(d\).

At still larger heights, however, the assumptions underlying the present model progressively cease to be appropriate. Heights comparable to the solar radius require a fully spherical treatment together with
the exact gravitational and Pannekoek--Rosseland potentials, whereas the present analysis assumes a plane-parallel geometry and the linear approximations of these potentials.

\section{Conclusions}
\label{sec:conclusions}

\subsection{Summary and discussions}
In this work we have investigated the combined effects of gravitational filtering, stochastic heating, and magnetic-field expansion in a collisionless plasma atmosphere representative of stellar atmospheres. The analysis has been carried out within a fully kinetic framework based on the Vlasov equation under the guiding-centre approximation, assuming particle motion along an expanding magnetic flux tube. This approach allowed us to derive fully analytical expressions for the density and temperature profiles and to identify the physical mechanisms governing their behaviour.

For a single-temperature boundary condition, we obtained analytical expressions for the density and temperature profiles in the presence of an expanding magnetic field. We showed that local charge neutrality implies that the plasma dynamics is governed by the combined action of gravity, the Pannekoek--Rosseland electric field, and magnetic-moment conservation. The latter progressively restricts the region of velocity space accessible to particles reaching a given height, producing a loss-cone distribution. As a consequence, the density is always reduced with respect to the corresponding unmagnetized atmosphere, while the plasma develops a pronounced temperature anisotropy, with an enhanced parallel temperature and a depleted perpendicular temperature.

More generally, we derived the asymptotic behaviour of the solution for an arbitrary magnetic field whose strength decreases monotonically with height faster than \(z^{-1}\). Under these assumptions, magnetic moment conservation introduces universal algebraic corrections to the gravitationally filtered atmosphere. In particular, the density acquires an additional depletion proportional to \(zB(z)\), while the perpendicular temperature exhibits the same asymptotic scaling and vanishes at large heights. In contrast, the parallel temperature returns asymptotically to the boundary value, although its detailed evolution at intermediate heights depends on the specific magnetic-field profile. The magnetic field present above a concentrated source, adopted throughout the numerical examples, represents a particular case of this more general class, recovering the scaling \(zB(z)\propto z^{-1}\). For the family of magnetic-field profiles given by Eq.~\eqref{eq:generalized_potential_field}, we show
that the parallel temperature develops a maximum as a result of the competition between magnetic moment conservation and gravity. We derive analytical scaling laws for the location and amplitude of this maximum and verify their accuracy through direct numerical solutions.

We then extended the formalism to multi-temperature boundary conditions generated by stochastic heating events. Despite the increased complexity of the distribution function, fully analytical
expressions for the density and temperature moments were again obtained. We showed that the main effects of magnetic-field expansion remain remarkably robust: the density is systematically reduced with
respect to the corresponding unmagnetized solution, while the plasma develops a pronounced anisotropy between the parallel and perpendicular temperature components.

Particular attention was devoted to the physically relevant regime of rare but intense heating events. In this limit, gravitational filtering progressively selects the hottest temperature components because of their larger gravitational scale heights. The combined
conservation of energy and magnetic moment then acts on these surviving populations by progressively restricting the region of velocity space that remains magnetically connected to the lower boundary, thereby producing a strong velocity-space anisotropy. As a consequence, the upper atmosphere becomes dominated by a hot and highly anisotropic particle population.

\subsection{Comparison with established fluid models and original gravitational filtering approaches}

We now compare the present kinetic description with standard fluid models of
coronal heating \cite{Klimchuk_2006,Klimchuk2015}. In these models, the
heating process enters the macroscopic energy equation as an energy input
within the coronal plasma, while the resulting thermodynamic structure is
determined by its interplay with processes such as thermal conduction and
radiative losses. Sustaining million-degree coronal temperatures therefore
requires a continuous balance between the energy supplied by heating and the
mechanisms that remove or redistribute this energy. In the present kinetic
framework, as well as in our previous works
\cite{Barbieri2023temperature,Barbieri2024b,Barbieri2025c}, the role of
stochastic heating is conceptually different. Heating is imposed at the lower
boundary through temporal fluctuations of the particle distribution rather
than as a volumetric heating term in a fluid energy equation. These
fluctuations generate a multi-temperature, non-Maxwellian boundary
distribution which, through gravitational filtering, preferentially populates
the upper atmosphere with more energetic particles. The million-degree
coronal temperature can therefore emerge from the collisionless selection of
the energetic particle populations injected at the lower boundary, rather
than from local heating of the plasma at coronal heights. Magnetic-field
expansion plays a distinct role: it does not provide an additional source of
energy but, through magnetic-moment conservation, modifies the distribution
of kinetic energy between the parallel and perpendicular degrees of freedom,
thereby generating the temperature anisotropy investigated in the present
work. The present model thus provides a complementary kinetic description in
which stochastic heating supplies the energetic particle populations,
gravitational filtering determines their increasing relative importance with
height, and magnetic-field expansion redistributes their kinetic energy
between the parallel and perpendicular degrees of freedom.

It is also worth emphasizing how the multi-temperature formulation adopted here and in previous works \cite{Barbieri2023temperature,Barbieri2024b,Barbieri2025,Barbieri2025b,Barbieri2025c} differs from original velocity-filtration model \cite{Scudder1992a,Scudder1992b}. In the original model of Scudder, suprathermal particles are prescribed a priori through a Kappa distribution at the lower boundary. Although gravitational filtering of this distribution produces a temperature inversion, the resulting temperature increases linearly with height and does not naturally reproduce a transition region followed by a hot corona. Finally, the original velocity-filtration model of Scudder does not include
the effects of magnetic-field expansion. Subsequent exospheric models
\cite{Pierrard1996,Maksimovic1997,Zouganelis2004} incorporated magnetic-field
expansion within the Liouville-mapping formalism, generally assuming
Kappa electron distributions at the exobase, and were developed to
describe the acceleration of the solar wind. Their lower boundary is located
at coronal altitudes, and therefore above the transition region. Among these
models, \cite{Zouganelis2004} places the exobase at comparatively low coronal
altitudes and provides a transonic solution describing the acceleration of
the solar wind from subsonic to supersonic velocities. The present model
addresses a different physical configuration: its lower boundary is placed at the base of the transition region, so that both the transition region and
the overlying hot corona are part of the atmospheric structure predicted by
the model. Moreover, we consider a closed magnetic geometry rather than an
open solar-wind configuration, and therefore no escaping wind population or
associated bulk outflow is present.

In contrast, in the present multi-temperature approach and in our previous works \cite{Barbieri2023temperature,Barbieri2024b,Barbieri2025,Barbieri2025b,Barbieri2025c} the non-thermal boundary distribution emerges from the coarse-graining of Maxwellian populations associated with stochastic heating events characterized by different temperatures. The resulting suprathermal component is therefore directly linked to the statistical properties of the heating process through the distribution \(\gamma(T)\) and the filling factor \(A_t\). This formulation also allows the dependence of the atmospheric structure on the heating statistics to be investigated; as shown in previous work, the resulting temperature inversion depends only weakly on the detailed functional form of \(\gamma(T)\). In the present study, this framework is further extended to an expanding magnetic field, allowing stochastic heating and gravitational filtering to be combined with magnetic-moment conservation and the resulting temperature anisotropy within a single kinetic description.

We also note that the representation of a non-Maxwellian distribution as a superposition of Maxwellian distributions with different temperatures was introduced theoretically in the framework of superstatistics \cite{BECKsuper}, where its general consequences for Hamiltonian systems were investigated. In the present framework, as well as in our previous works discussed above, this multi-temperature distribution is not postulated as an equilibrium distribution associated with fluctuations of the background temperature. Rather, it emerges from the coarse-graining procedure and describes the statistical superposition of particle populations produced by different heating events at the base of the transition region. This provides a direct physical interpretation of the temperature distribution $\gamma(T)$ in terms of the statistics of the heating process. We further note that the standard Kappa distribution can be recovered as a particular case of this more general superstatistical representation for an appropriate choice of $\gamma(T)$ \cite{lazar2021kappa}.

Interestingly, the density and temperature profiles are only weakly sensitive to the detailed functional form of \(\gamma(T)\), indicating that the gravitational-filtering mechanism is robust with respect to the specific statistics of the heating events. At the same time, \(\gamma(T)\) retains a clear physical interpretation, as it describes the statistical distribution of the temperatures generated by stochastic heating at the lower boundary. The multi-temperature formulation therefore provides a direct connection between  the statistical properties of the underlying heating process and the resulting non-Maxwellian boundary distribution.

\subsection{Future perspectives}

Within the guiding-centre approximation adopted here, these properties depend on the magnetic-field expansion through the loss-cone parameter, Eq.~\eqref{def_C(z)}, rather than on the absolute magnetic-field strength. The present description therefore remains applicable as long as the first adiabatic invariant is conserved, namely when the particle Larmor radius remains much smaller than the characteristic scale over which the magnetic field varies.

Such strongly anisotropic velocity distributions are expected to be susceptible to kinetic instabilities and to pitch-angle scattering, which would tend to relax the anisotropy and repopulate the loss cone. Likewise, even weak Coulomb collisions would produce a similar effect. The interplay between magnetic-gravity filtration and these relaxation mechanisms lies beyond the scope of the present work and will be investigated in a future study.

Overall, the present work demonstrates that gravitational filtering and magnetic moment conservation play complementary roles in determining the structure of collisionless stellar atmospheres. Gravitational filtering controls which temperature components populate the upper atmosphere, whereas magnetic moment conservation governs the distribution of their kinetic energy between the parallel and perpendicular degrees of freedom through the formation of a loss cone. Together, these mechanisms provide a simple, fully analytical, and self-consistent kinetic framework for understanding the formation of non-trivial density and temperature profiles in magnetized plasmas. The present model represents the collisionless limit of weakly collisional magnetized atmospheres, providing a natural reference against which the effects of collisions and pitch-angle scattering can be assessed.

Finally, we note that the interest of the present model extends beyond its specific application to stellar atmospheres. From a more general perspective, the solution provides a detailed analytical description of the kinetic dynamics of a collisionless two-component plasma subject to the same external force acting on both species and embedded in a non-uniform magnetic field. In this sense, the results presented here may be viewed as a fundamental reference solution for the study of collisionless plasmas in expanding magnetic geometries.

Concerning the solar application, the present analytical solution should not be regarded as a fully realistic description of the low corona. The adopted magnetic-field configuration is intended to represent the nearly vertical legs of a sufficiently large coronal loop, for which the field direction varies only weakly over the spatial region considered here. For smaller magnetic structures associated with localized heating events, however, the curvature of the field lines may become important, and a dipolar or loop-like magnetic geometry would provide a more realistic description \cite{Demoulin_1994}. In such a geometry, both the gravitational potential along the field line and the magnetic-field strength would vary with the curvilinear coordinate, thereby modifying the particle trajectories and the corresponding accessible regions of phase space. Extending the present treatment to curved magnetic geometries therefore represents an important direction for
future work.

In addition, Coulomb collisions are expected to play an important role in the dense plasma of the transition region \citep{Landi-Pantellini2001}. Collisions tend to isotropize the velocity distribution and to repopulate the trapped region of phase space that is excluded in the collisionless limit by the presence of the loss cone. The competition between magnetic
moment conservation and collisional isotropization therefore represents a key ingredient for understanding the transition from collisionless to weakly collisional regimes in expanding stellar atmospheres. Turbulent pitch-angle scattering may produce an additional redistribution of particles in velocity space and may consequently also modify the anisotropies predicted by the collisionless model. The importance of collisional and turbulent scattering for electron transport in an inhomogeneous magnetic field has also been investigated by
\citet{Tang_2023}.

The present collisionless solution should therefore be regarded as a useful analytical benchmark against which more sophisticated transport models can be compared. In future work, it would be worthwhile to extend the collisional formalism developed by \citet{barbieri2026}, which was restricted to the case of a uniform magnetic field, to the more general magnetic geometries considered here, and subsequently to curved loop-like configurations. Such an extension will make it possible to investigate how collisions and turbulent scattering modify the different particle populations generated by stochastic heating, to determine which populations remain anisotropic and which become effectively isotropized, and to quantify the progressive repopulation of the trapped region of velocity space. In particular, it will be possible to study how the efficiency of scattering depends on the temperature of the various populations, thereby identifying which components preserve the collisionless signatures described in the present work and which are instead driven toward more isotropic
distributions.

A direct quantitative comparison with observed temperature anisotropies would therefore be premature at this stage, since the present collisionless model does not account for the collisional relaxation that acts to reduce the anisotropy generated by magnetic-field expansion. Once the interplay between magnetic-field expansion and collisional isotropization is consistently characterized, the magnitude of the resulting anisotropy can be more realistically quantified and meaningfully compared with observations.


\begin{acknowledgments}
The authors wish to thank the anonymous reviewers for the
valuable comments and suggestions that helped improve the clarity and quality of this work.
L.B. wants to thank Sorbonne Universit\'e for financial support within the framework of the Initiative Physique des Infinis. L.B. wants to thank Pierfrancesco Di Cintio for useful discussions. D.V. is supported by STFC Consolidated Grant ST/W001004/1.
\end{acknowledgments}

\section*{Conflict of interest}
The author has no conflicts to disclose.

\section*{DATA AVAILABILITY}
The data that support the findings of this study are available from
the corresponding author upon reasonable request.

\bibliography{biblio}

\appendix

\section{Kinetic moments}\label{appendix:moments}
\subsection{Main equations}  \label{sec:Main_equations}

The density is defined as
\begin{equation}
\label{densitymom}
    n_\alpha(z)
    =
    \int d^3v\,
    f_\alpha(z,\mathbf v).
\end{equation}
The parallel and perpendicular second moments are
\begin{equation}
    \langle v_{\parallel,\alpha}^2\rangle
    =
    \int d^3v\,
    v_\parallel^2\,
    f_\alpha(z,\mathbf v),
\end{equation}
\begin{equation}
    \langle v_{\perp,\alpha}^2\rangle
    =
    \int d^3v\,
    v_{\perp,x}^2\,
    f_\alpha(z,\mathbf v),
\end{equation}
   where
\begin{equation}
    v_\parallel=v\cos\theta,
    \qquad
    v_{\perp,x}=v\sin\theta\cos\varphi.
\end{equation}
    The corresponding kinetic temperatures are
\begin{equation}
\label{paralleltempmom}
    T_{\parallel,\alpha}(z)
    =
    \frac{m_\alpha}{k_B}
    \frac{
    \langle v_{\parallel,\alpha}^2\rangle
    }{
    n_\alpha(z)
    },
\end{equation}
\begin{equation}
\label{perptempmom}
    T_{\perp,\alpha}(z)
    =
    \frac{m_\alpha}{k_B}
    \frac{
    \langle v_{\perp,\alpha}^2\rangle
    }{
    n_\alpha(z)
    },
\end{equation}
   while the total temperature is
\begin{equation}
\label{totaltempmom}
    T_\alpha(z)
    =
    \frac{
    T_{\parallel,\alpha}(z)
    +
    2\,T_{\perp,\alpha}(z)
    }{3}.
\end{equation}

\subsection{Evaluation of the angular integrals}
\label{sec:Evaluation_angular}

Since the kinetic energy depends only on the magnitude of the velocity, it is spherically symmetric in velocity space. We therefore adopt spherical velocity coordinates with the polar axis oriented along
\(\mathbf{B}\),
\begin{equation}
d^3v
=
v^2\sin\theta
\,dv\,d\theta\,d\varphi.
\end{equation}

The parallel moments are evaluated over the populated region of velocity space. For \(v<v_{B,\alpha}\), all pitch angles are populated, whereas for \(v>v_{B,\alpha}\) only two symmetric polar caps remain
magnetically connected to the lower boundary. Therefore,
\begin{equation}
\begin{aligned}
\langle v_{\parallel,\alpha}^k\rangle
={}&
2\pi
\Bigg[
\int_0^{v_{B,\alpha}}
dv\,
v^{k+2}
f_\alpha(z,\mathbf v)
\\
&\qquad\times
\int_0^\pi
d\theta\,
\sin\theta\,
\cos^k\theta
\\
&+
\int_{v_{B,\alpha}}^{+\infty}
dv\,
v^{k+2}
f_\alpha(z,\mathbf v)
\\
&\qquad\times
\left(
\int_0^{\theta_m}
+
\int_{\pi-\theta_m}^{\pi}
\right)
d\theta
\\
&\qquad\qquad\times
\sin\theta\,
\cos^k\theta
\Bigg].
\end{aligned}
\end{equation}

For the first angular integral one finds
\begin{equation}
\int_0^\pi
d\theta\,
\sin\theta
\cos^k\theta
=
\frac{1+(-1)^k}{k+1},
\end{equation}
whereas over the populated polar caps
\begin{align}
&
\int_0^{\theta_m}
d\theta\,
\sin\theta
\cos^k\theta
+
\int_{\pi-\theta_m}^{\pi}
d\theta\,
\sin\theta
\cos^k\theta
\nonumber\\
&\qquad=
\frac{1+(-1)^k}{k+1}
\left(
1-\cos^{k+1}\theta_m
\right).
\end{align}

Using Eq.~\eqref{def_cos_theta_m}, the parallel moments become
\begin{equation}
\label{parallelmom}
\begin{aligned}
\langle v_{\parallel,\alpha}^k\rangle
={}&
\frac{2\pi}{k+1}
\left[
1+(-1)^k
\right]
\Bigg\{
\\
&
\int_0^{+\infty}
dv\,
v^{k+2}
f_\alpha(z,\mathbf v)
\\
&
-
\mathcal{C}^{k+1}(z)
\int_{v_{B,\alpha}}^{+\infty}
dv\,
v
\\
&\qquad\times
\left(
v^2-v_{B,\alpha}^2
\right)^{(k+1)/2}
\\
&\qquad\times
f_\alpha(z,\mathbf v)
\Bigg\}.
\end{aligned}
\end{equation}
The odd moments vanish because the populated regions in the upper and lower hemispheres are symmetric with respect to \(v_\parallel=0\).

For the perpendicular second moment, we consider the \(x\)-component of the particle velocity, with the \(z\)-axis chosen along the magnetic
field,
\begin{equation}
v_{x,\alpha}
=
v\sin\theta\cos\varphi.
\end{equation}

Because the system is axisymmetric around \(\mathbf{B}\), the two perpendicular directions are equivalent. Therefore $\langle v_{x,\alpha}^2\rangle = \langle v_{y,\alpha}^2\rangle$. The second moment of the \(x\)-component must again be evaluated
separately over the fully populated and loss-cone-restricted regions of velocity space. For \(v<v_{B,\alpha}\), the complete angular domain is
populated, whereas for \(v>v_{B,\alpha}\) only the two symmetric polar caps remain connected to the lower boundary. Hence,
\begin{equation}
\begin{aligned}
\langle v_{x,\alpha}^2\rangle
={}&
\int_0^{v_{B,\alpha}}
dv\,
v^4 f_\alpha(z,v)
\\
&\times
\int_0^{2\pi}
d\varphi\,
\cos^2\varphi
\int_0^\pi
d\theta\,
\sin^3\theta
\\
&+
2
\int_{v_{B,\alpha}}^{+\infty}
dv\,
v^4 f_\alpha(z,v)
\\
&\times
\int_0^{2\pi}
d\varphi\,
\cos^2\varphi
\\
&\times
\left(
\int_0^{\theta_m}
+
\int_{\pi-\theta_m}^{\pi}
\right)
d\theta\,
\sin^3\theta .
\end{aligned}
\end{equation}
where we use 
\begin{equation} \nonumber
\int_{\pi-\theta_m}^{\pi}
d\theta\,
\sin^3\theta
=
\int_0^{\theta_m}
d\theta\,
\sin^3\theta .
\end{equation}

The azimuthal integral is
\begin{equation}
\int_0^{2\pi}
d\varphi\,
\cos^2\varphi
=
\pi,
\end{equation}
while
\begin{equation}
\int_0^{\theta_m}
d\theta\,
\sin^3\theta
=
\frac{2}{3}
-
\cos\theta_m
+
\frac{1}{3}\cos^3\theta_m.
\end{equation}
Therefore,
\begin{equation}
\begin{aligned}
\langle v_{x,\alpha}^2\rangle
={}&
\pi
\Bigg[
\frac{4}{3}
\int_0^{v_{B,\alpha}}
dv\,
v^4 f_\alpha(z,v)
\\
&+
\int_{v_{B,\alpha}}^{+\infty}
dv\,
v^4
\\
&\qquad\times
f_\alpha(z,v)
\Bigg(
\frac{4}{3}
-
2\cos\theta_m
\\
&\qquad\qquad
+
\frac{2}{3}\cos^3\theta_m
\Bigg)
\Bigg].
\end{aligned}
\end{equation}

Using Eq.~\eqref{def_cos_theta_m} and combining the two radial integration domains, one obtains
\begin{equation}
\label{perpmom}
\begin{aligned}
\langle v_{x,\alpha}^2\rangle
={}&
\pi
\Bigg[
\frac{4}{3}
\int_0^{+\infty}
dv\,
v^4
f_\alpha(z,v)
\\
&-
2\,\mathcal{C}(z)
\int_{v_{B,\alpha}}^{+\infty}
dv\,
v^3
\\
&\qquad\times
\sqrt{v^2-v_{B,\alpha}^2}\,
f_\alpha(z,v)
\\
&+
\frac{2}{3}
\mathcal{C}^3(z)
\int_{v_{B,\alpha}}^{+\infty}
dv\,
v
\\
&\qquad\times
\left(
v^2-v_{B,\alpha}^2
\right)^{3/2}
f_\alpha(z,v)
\Bigg].
\end{aligned}
\end{equation}

By axisymmetry around the magnetic-field direction, the same result holds for the \(y\)-component. Accordingly, the total perpendicular second moment is
\begin{equation}
\langle v_{\perp,\alpha}^2\rangle
=
\langle v_{x,\alpha}^2\rangle
+
\langle v_{y,\alpha}^2\rangle
=
2\langle v_{x,\alpha}^2\rangle.
\end{equation}

\subsection{Velocity integrals}
\label{sec:Velocity_integrals}

Introducing
\begin{equation}
    \beta_\alpha
    =
    \frac{m_\alpha}{2\,k_B T}.
\end{equation}
The velocity integrals required for the density and pressure
moments are
\begin{equation}
    \int_0^{+\infty}
    v^2\, e^{-\beta_\alpha v^2}\,dv
    =
    \frac{\sqrt{\pi}}{4}
    \beta_\alpha^{-3/2},
\end{equation}
\begin{equation}
    \int_0^{+\infty}
    v^4\, e^{-\beta_\alpha v^2}\,dv
    =
    \frac{3\sqrt{\pi}}{8}
    \beta_\alpha^{-5/2},
\end{equation}
\begin{equation}
\begin{aligned}
&
\int_{v_{B,\alpha}}^{+\infty}
v\,
\sqrt{v^2-v_{B,\alpha}^2}\,
e^{-\beta_\alpha v^2}\,dv
\\
&\qquad=
\frac{\sqrt{\pi}}{4}\,
\beta_\alpha^{-3/2}
e^{-\as(z,T)}.
\end{aligned}
\end{equation}

\begin{equation}
\begin{aligned}
&
\int_{v_{B,\alpha}}^{+\infty}
v\,
\left(
v^2-v_{B,\alpha}^2
\right)^{3/2}
e^{-\beta_\alpha v^2}\,dv
\\
&\qquad=
\frac{3\sqrt{\pi}}{8}\,
\beta_\alpha^{-5/2}
e^{-\as(z,T)}.
\end{aligned}
\end{equation}
   and
\begin{align}
&
\int_{v_{B,\alpha}}^{+\infty}
v^3
\sqrt{
v^2-v_{B,\alpha}^2
}
\,e^{-\beta_\alpha v^2}\,dv
\nonumber\\
&=
\frac{\sqrt{\pi}}{8}
\beta_\alpha^{-5/2}
\left[
3+2\as(z,T)
\right]
e^{-\as(z,T)}.
\end{align}

Substituting the above expressions into Eqs.~\eqref{parallelmom} and \eqref{perpmom} provides the density, parallel pressure, and perpendicular pressure associated with a single Maxwellian component reported in Section \ref{sec:single-temperature}.
The kinetic moments of the multi-temperature distribution are obtained by integrating the corresponding single-temperature contributions over the heating-temperature distribution $\gamma(T)$.
The resulting expressions for the density, parallel temperature, perpendicular temperature, and scalar temperature are reported in Section \ref{sec:multi-temperature}.

\section{Sign of the derivative of \(F(\phis)\)}
\label{appendix:derivative}

In this Appendix, we show that the function \(F(\phis)\) defined by Eq.~\eqref{F} is strictly decreasing. To simplify the notation, we introduce
\begin{equation}
\label{defG}
\begin{aligned}
G(V_\alpha)
={}&
\exp\!\left(
-\frac{V_\alpha}{k_B\Tb}
\right)
\\
&\times
\Biggl[
1
-
\mathcal{C}(z)
\\
&\qquad\times
\exp\!\left(
-\frac{V_\alpha}{k_B\Tb}
\left(
1-\frac{1}{\mathcal{C}^2(z)}
\right)
\right)
\Biggr].
\end{aligned}
\end{equation}
The function \(F\) can then be written as
\begin{equation}
F(\phis)
=
G\!\left(V_p\right)
-
G\!\left(V_e\right),
\end{equation}

Differentiating with respect to \(\phis\), we obtain
\begin{equation}
\frac{dF}{d\phis}
=
e\,\frac{\partial G}{\partial V_\alpha}(V_p)
-e\,\frac{\partial G}{\partial V_\alpha}(V_e).
\label{eq:Fprimecompact}
\end{equation}
The derivative of \(G\) is
\begin{equation}
\label{Gprime}
\begin{aligned}
\frac{\partial G}{\partial V_\alpha}
={}&
-\frac{1}{k_B\Tb}
\exp\!\left(
-\frac{V_\alpha}{k_B\Tb}
\right)
\\
&\times
\Bigg[
1
-
\mathcal{C}(z)
\left(
2-\frac{1}{\mathcal{C}^2(z)}
\right)
\\
&\qquad\times
\exp\!\left(
-\frac{V_\alpha}{k_B\Tb}
\left(
1-\frac{1}{\mathcal{C}^2(z)}
\right)
\right)
\Bigg].
\end{aligned}
\end{equation}

For \(z>0\) and finite, the expansion of the magnetic flux tube implies
\begin{equation}
0<\mathcal{C}(z)<1.
\end{equation}
Then 
\begin{equation}
2-\frac{1}{\mathcal{C}^2(z)} <1
\end{equation}
Moreover, positivity of the density requires
\begin{equation}
\mathcal{C}(z)
\exp\!\left(
-\frac{V_\alpha}{k_B\Tb}
\left(
1-\frac{1}{\mathcal{C}^2(z)}
\right)
\right)
\leq 1.
\label{eq:densityconstraint}
\end{equation}
It follows that the term in between square brackets in Eq.~\eqref{Gprime} is positive. Therefore
\begin{equation}
\frac{\partial G}{\partial V_\alpha}<0.
\end{equation}

Since both terms on the right-hand side of
Eq.~\eqref{eq:Fprimecompact} are negative, it follows that

\begin{equation}
\frac{dF}{d\phis}<0
\end{equation}
throughout the domain. In particular, at \(\phis=0\), \(V_{e}=V_{p}=m\,g\,z\) and
\begin{equation}
\left.
\frac{dF}{d\phis}
\right|_{\phis=0}
=
2e\,\frac{\partial G}{\partial V_\alpha}(m\,g\,z)
<0.
\end{equation}
Hence \(F(\phis)\) is strictly decreasing and can possess at most one
zero.

\section{Uniqueness of the electrostatic potential in the multi-temperature case}
\label{appendix:multitemp_uniqueness}

In this Appendix we show that the local charge-neutrality condition admits a unique solution also in the multi-temperature case. Using the density derived in Section~\ref{sec:multi-temperature}, the neutrality condition can be written as
\begin{equation}
F_T(\phis)=0,
\end{equation}
where
\begin{align}
F_T(\phis)
=
&
A_t
\int_{T_0}^{+\infty}
dT\,\gamma(T)
\left[
G(V_p,T)
-
G(V_e,T)
\right]
\nonumber\\
&
+
(1-A_t)
\left[
G(V_p,T_0)
-
G(V_e,T_0)
\right],
\label{eq:Fmulti}
\end{align}
with \(G(V_{\alpha},T)\) defined by Eq.~\eqref{defG} after replacing \(T_b\) by \(T\), while  \(G(V_{\alpha},T_0)\) is obtained from the same expression by replacing \(T_b\) by \(T_0\). 

It is immediately verified that
\begin{equation}
F_T(0)=0,
\end{equation}
so that \(\phis=0\) is a solution. To establish uniqueness, it is sufficient to prove that \(F_T(\phis)\) is strictly monotonic. Differentiating Eq.~\eqref{eq:Fmulti} with respect to \(\phis\) gives
\begin{align}
\frac{dF_T}{d\phis}
=
&
\,e\,A_t
\int_{T_0}^{+\infty}
dT\,\gamma(T)
\left[
\frac{\partial G}{\partial V_\alpha}(V_p,T)
+
\frac{\partial G}{\partial V_\alpha}(V_e,T)
\right]
\nonumber\\
&
+
e\,(1-A_t)
\left[
\frac{\partial G}{\partial V_\alpha}(V_p,T_0)
+
\frac{\partial G}{\partial V_\alpha}(V_e,T_0)
\right].
\label{eq:Fmulti_prime}
\end{align}

The derivative \(G'(V_\alpha,T)\) has exactly the same form as in the single-temperature case discussed in Appendix~\ref{appendix:derivative}, i.e. given by Eq. \eqref{Gprime}. Exactly the same arguments used in Appendix~\ref{appendix:derivative} show that
\begin{equation}
\frac{\partial G}{\partial V_\alpha}(V_\alpha,T)<0,
\end{equation}
for every temperature \(T\), every admissible value of \(V_\alpha\), and every finite \(z>0\). 

Since the temperature distribution satisfies
\begin{equation}
\gamma(T) > 0,
\end{equation}
the integral appearing in Eq.~\eqref{eq:Fmulti_prime} is a weighted average of strictly negative functions and is therefore itself strictly negative. The contribution associated with the background Maxwellian at temperature \(T_0\) is also strictly negative. Then, we conclude that
\begin{equation}
\frac{dF_T}{d\phis}<0,
\end{equation}
for every value of \(\phis\). Hence \(F_T(\phis)\) is strictly decreasing and can therefore possess at most one zero. Therefore, local charge neutrality uniquely determines a vanishing supplementary electrostatic potential also in the multi-temperature case, independently of the particular form of the boundary temperature distribution \(\gamma(T)\).

\end{document}